\documentstyle[amssymb,prb,aps,epsfig]{revtex}
\begin{document}
\draft
\title{Coupled mesoscopic superconductors}
\author{B. J. Baelus, S. V. Yampolskii\cite{Yampolskii} and F. M. Peeters\cite
{peeters}}
\address{Departement Natuurkunde, Universiteit Antwerpen (UIA), Universiteitsplein 1,%
\\
B-2610 Antwerpen, Belgium}
\date{\today}
\maketitle

\begin{abstract}
The magnetic coupling between two concentric mesoscopic superconductors with
non-zero thickness is studied using the nonlinear Ginzburg-Landau theory. We
calculated the free energy, the expelled field, the total field profile, the
Cooper-pair density and the current density distribution. By putting a
smaller superconducting disk or ring in the center of a larger ring, the
properties change drastically. Extra ground state transitions are found,
where the total vorticity stays the same, but the vorticity of the inner
superconductor changes by one. Due to the magnetic coupling, the current in
the external ring exhibits extra jumps at the transition fields where the
vorticity of the inner superconductor changes. In this case, for certain
temperatures, re-entrant behavior and switching on and off of the
superconducting behavior of the rings are found as function of the magnetic
field. A $H-T$ phase diagram is obtained for the situation where the inner
ring has a higher critical temperature than the outer ring. An analytic
expression for the magnetic coupling is obtained for thin rings and extreme
type-II superconductors.
\end{abstract}


\section{Introduction}

Single mesoscopic superconducting rings~\cite
{Boek,Vloeberghs,Berger,Moshchalkov,Horane,Fomin,Bruyndoncx1,PRB61,PRB63,EffRad,Lindelof}
and disks~\cite
{Buisson,Benoist,Geim1,PRL79,Geim2,PRB57,PRL81,PhysC2000,Palacios,PRL83,PRB62,Akkermans,Geim3,charge}
have attracted a lot of attention during the last years. The superconducting
properties of these single samples have been studied experimentally and are
rather well described and understood theoretically.

Many of these studies were limited to disks~\cite
{Buisson,Benoist,Palacios,PRL83,PRB62,Akkermans} and rings~\cite
{Bruyndoncx1,PRB63} of zero thickness. In such a case the magnetic field
induced by the supercurrents are neglected and one assumes that the total
field equals the external field, which is taken uniform. In the present
paper disks and rings with {\it finite} thickness are considered. In this
case the superconductor tries to expell the field. Supercurrents are induced
creating a local magnetic field opposite to the external one. This leads to
a decrease of magnetic field inside the superconductor and an increase near
the sample edges. For sufficiently large disks and magnetic fields, the
magnetic field penetrates the superconductor and vortices are created, which
corresponds to regions of local field compression. In small disks, only
circular symmetric vortex states nucleate in the center, which are called
giant vortex states. Encircling the vortex on a closed loop, the phase
change of the order parameter equals $L$ times $2\pi $, where $L$ is the
angular momentum or vorticity. In larger disks, seperate vortices can
nucleate creating a multivortex state and the vorticity $L$ is just the
number of vortices. In the case of rings, the magnetic field can be expelled
to the inside or to the outside of the ring or partially to both sides.

In the present paper we want to understand what will happen if two
mesoscopic superconductors are put close to each other but are electrically
isolated. When an uniform magnetic field is applied, it will be locally
altered by each of the superconductors. In some regions of space the field
will be expelled from the superconducting disk or ring, while in other
regions it will be compressed into vortices penetrating the sample or
compressed into the inside of a superconducting ring. This results in a
strongly nonuniform total field, which is the superposition of the applied
field and the field created by the supercurrents. Another superconductor
will interact with this nonuniform field. In this way the superconductors
are coupled by the magnetic field and they are interacting with each other.
This coupling will influence the properties of the two superconductors. It
is this coupling that will be studied in the present paper.

The magnetic coupling between normal mesoscopic rings is well known. Wang 
{\it et al}~\cite{Wang} studied theoretically the persistent currents of $N$
normal rings placed periodically on the same plane. Because of the mutual
inductance between two rings, the electric current in one ring produces an
induced flux in the other ring, creating an extra current in this ring. They
found that the mutual inductance between normal rings enhances the
persistent current as was the case in the experiments described in Ref.~\cite
{Levy}.

Correlations in arrays of magnetically coupled superconducting aluminium
rings were investigated experimentally by Davidovi\'{c} {\it et al}~\cite
{Davidovic}. They used ultrasensitive susceptibility techniques and scanning
Hall probe microscopy to study arrays of electrically isolated
superconducting rings of micron-size. When the external flux is close to $%
\Phi _{0}/2$,\ the magnetic moments produced by the supercurrents in such
rings are analogous to Ising spins. Magnetic moments parallel to the applied
magnetic field can be called ''spin up'', while those in the opposite
direction ''spin down''. Via their dipolar magnetic fields, neighbouring
rings can interact antiferromagnetically and the different rings are
influencing each other.

Recently, Morelle {\it et al}~\cite{Morelle} studied experimentally the
magnetic interaction between two superconducting mesoscopic aluminium rings,
close to the superconducting/normal phase transition. In their sample, a
smaller ring was placed in the center of the larger ring. Using resistivity
measurements the phase boundary was obtained for the two-ring structure as
well as for the reference single ring. In both systems, Little-Parks
oscillations were observed in the $H-T$ phase diagram. The modification of
the $T_{c}(H)$ oscillations of the outer ring is seen in the Fourier
spectrum of the $T_{c}(H)$ line due to the coupling between the outer and
the inner ring. They suggested that an inner ring made from a different
superconductor with a higher critical temperature would increase the
magnetic coupling between the two rings.

In the present paper we present a theoretical investigation of the
properties of two coupled mesoscopic superconductors. Our main attention
will go to the interaction between the two superconductors. How are they
influencing each other? How do the superconducting properties of a single
ring change when another superconductor is placed in its center? Therefore,
we consider two different configurations: (i) a ring-disk configuration
where a small disk is placed in the center of a larger ring, and (ii) a
ring-ring configuration where a small ring is placed in the center of the
larger ring as in the experiment of Ref.~\cite{Morelle}. We will also give
an example of a ring-ring system where the inner ring is made from a
different superconductor with a higher critical temperature. Our theoretical
analysis is based on a full self-consistent numerical solution of the
coupled non-linear Ginzburg-Landau equations. Since we consider sufficiently
narrow rings and small disks, only axial symmetric giant vortex states will
nucleate~\cite{PRB61}. Therefore, the equations can be solved for a fixed
value of the vorticity. The magnetic field profile outside and inside the
superconductor is obtained self-consistently and, therefore, the full
demagnetization effect is included in our approach.

The paper is organized as follows: In Sec.~II we describe the theoretical
formalism. Our results for the ring-disk configuration are discussed in
Sec.~III. In Sec.~IV the ring-ring configuration is studied. In Sec.~V we
calculate the $H-T$ phase diagram for the two ring system, where the inner
ring is made of a different material with a higher critical temperature. In
Sec.~VI we analytically calculate the energy of two coupled {\it thin} rings
made of a different material for high values of $\kappa $. Finally, in Sec.
VII our results are summarized.

\section{Theoretical formalism}

\begin{figure}[tbp]
\begin{center}
\epsfig{file=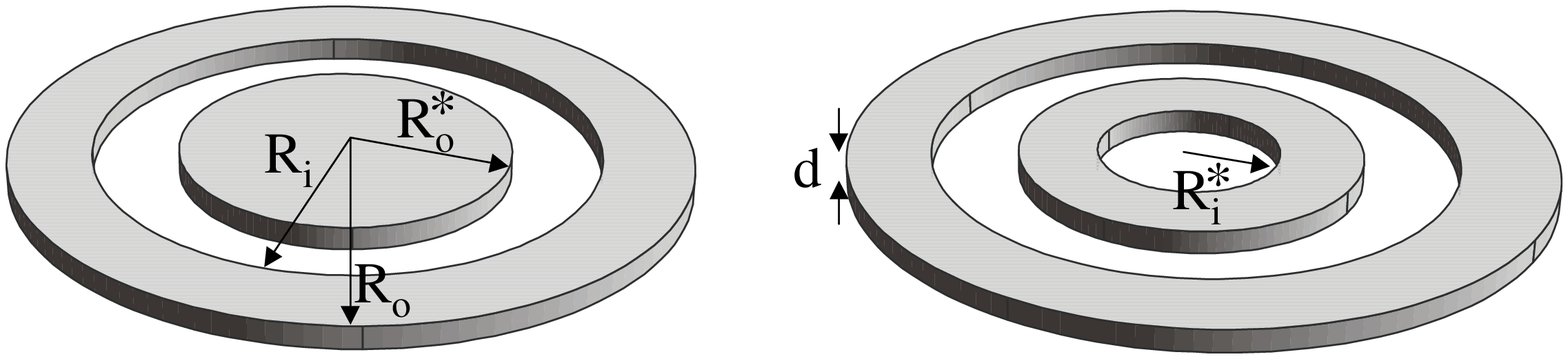, width=150mm, height=110mm, clip=}
\end{center}
\caption{Schematical outline of the considered configurations; the ring-disk
configuration (left) and the ring-ring configuration (right).}
\label{samples}
\end{figure}

We consider a superconducting ring with inner radius $R_{i}$, outer radius $%
R_{o}$ and thickness $d$ immersed in a insulating medium (for example
vacuum). In the center of this ring a superconducting disk with radius $%
R_{o}^{\ast }$ or another superconducting ring with inner radius $%
R_{i}^{\ast }$ and outer radius $R_{o}^{\ast }$ is placed with the same
thickness (see Fig.~\ref{samples}). The whole sample is placed in a
perpendicular uniform magnetic field $\overrightarrow{H}=\left(
0,0,H_{0}\right) $. To solve this problem, we expand our previous approach
for thin superconducting disks~\cite{PRB57} to a system of two axial
symmetric superconductors each made of a\ different material. In the present
paper we solve the system of two coupled nonlinear Ginzburg-Landau equations
which determine the distribution of both the superconducting order parameter 
$\Psi \left( \overrightarrow{r}\right) $ and the vector potential $%
\overrightarrow{A}\left( \overrightarrow{r}\right) $ inside and outside the
superconductor, 
\begin{equation}
\frac{1}{2m}\left( -i\hbar \overrightarrow{\nabla }-\frac{2e\overrightarrow{A%
}}{c}\right) ^{2}\Psi =-\alpha _{i,o}\Psi -\beta _{i,o}\Psi \left| \Psi
\right| ^{2}\text{ ,}  \eqnum{1a}  \label{GL1}
\end{equation}
\begin{equation}
\overrightarrow{\nabla }\times \overrightarrow{\nabla }\times 
\overrightarrow{A}=\frac{4\pi }{c}\overrightarrow{j}\text{ ,}  \eqnum{1b}
\label{GL2}
\end{equation}
where for the inner superconductor the parameters are $\alpha _{i,o}=\alpha
_{i}$, $\beta _{i,o}=\beta _{i}$, and for the outer ring $\alpha
_{i,o}=\alpha _{o}$, $\beta _{i,o}=\beta _{o}$. The density of
superconducting current $\overrightarrow{j}$ is given by 
\begin{equation}
\overrightarrow{j}=\frac{e\hbar }{im}\left( \Psi ^{\ast }\overrightarrow{%
\nabla }\Psi -\Psi \overrightarrow{\nabla }\Psi ^{\ast }\right) -\frac{4e^{2}%
}{mc}\left| \Psi \right| ^{2}\overrightarrow{A}\text{ ,}  \eqnum{2}
\label{GL3}
\end{equation}
and $m$ is the mass of the Cooper-pair. Since we only consider circular
symmetric rings and disks, we use cylindrical coordinates. Any position in
space will be expressed by $\overrightarrow{r}=\left( \rho ,\phi ,z\right) ,$
where $\rho $ is the radial distance from the center, $\phi $ is the
azimuthal angle and $z$ is the perpendicular direction. The sample lies
between $z=-d/2$ and $z=d/2$.

Eqs.~(\ref{GL1})-(\ref{GL3}) have to be supplemented by boundary conditions
for $\Psi \left( \overrightarrow{r}\right) $ and $\overrightarrow{A}\left( 
\overrightarrow{r}\right) $. The boundary condition for the superconducting
condensate at the superconductor/insulator surfaces is given by 
\begin{equation}
\left. \overrightarrow{n}\cdot \left( -i\hbar \overrightarrow{\nabla }-\frac{%
2e\overrightarrow{A}}{c}\right) \right| _{\rho =R_{1}^{\ast },R_{2}^{\ast
},R_{1},R_{2}}\Psi =0\text{ ,}  \eqnum{3}  \label{BC1}
\end{equation}
where $\overrightarrow{n}$ is the unit vector in the radial direction. This
condition expresses that the superconducting current in the radial direction
vanishes at the sample surface. The boundary for the vector potential has to
be taken far away from the disk 
\begin{equation}
\left. \overrightarrow{A}\right| _{\overrightarrow{r}\rightarrow \infty }=%
\frac{1}{2}\overrightarrow{e}_{\phi }H_{0}\rho \text{ ,}  \eqnum{4}
\label{BC2}
\end{equation}
where the field equals the applied magnetic field $\overrightarrow{H}=\left(
0,0,H_{0}\right) $.

To rewrite Eqs.~(\ref{GL1})-(\ref{GL3}) in dimensionless variables, we
express all distances in units of the coherence length of the outer
superconductor $\xi _{o}=\hbar /\sqrt{-2m\alpha _{o}}$, the order parameter
in $\Psi _{0,o}=\sqrt{-\alpha _{o}/\beta _{o}}$, the vector potential in $%
c\hbar /2e\xi _{o}$ and the magnetic field in $H_{c2,o}=c\hbar /2e\xi
_{o}^{2}=\kappa _{o}\sqrt{2}H_{c,o},$ where $H_{c,o}=\sqrt{4\pi \alpha
_{o}^{2}/\beta _{o}}$ is the thermodynamical critical field of the outer
superconductor and $\kappa _{o}=\lambda _{o}/\xi _{o}$ is the
Ginzburg-Landau parameter for this superconductor. The penetration depth of
the outer ring is given by $\lambda _{o}=\sqrt{m/\pi }\left( c/4e\right)
/\Psi _{0,o}$. Using these dimensionless variables and the London gauge, div 
$\overrightarrow{A}$ = 0, we can rewrite Eqs.~(\ref{GL1})-(\ref{GL3}) in the
following form: 
\begin{equation}
\left( -i\overrightarrow{\nabla }-\overrightarrow{A}\right) ^{2}\Psi =\Psi
\left( \frac{\xi _{o}^{2}}{\xi _{i,o}^{2}}-\frac{\kappa _{i,o}^{2}}{\kappa
_{o}^{2}}\left| \Psi \right| ^{2}\right) \text{ ,}  \eqnum{5a}  \label{GLd1}
\end{equation}
\begin{equation}
-\kappa _{o}^{2}\nabla ^{2}\overrightarrow{A}=\frac{1}{2i}\left( \Psi ^{\ast
}\overrightarrow{\nabla }\Psi -\Psi \overrightarrow{\nabla }\Psi ^{\ast
}\right) -\left| \Psi \right| ^{2}\overrightarrow{A}\text{ ,}  \eqnum{5b}
\label{GLd2}
\end{equation}
where $\xi _{i,o}=\xi _{i}$, $\kappa _{i,o}=\kappa _{i}$ in the inner
superconductor and $\xi _{i,o}=\xi _{o}$, $\kappa _{i,o}=\kappa _{o}$ in the
outer ring. The first boundary condition becomes 
\begin{equation}
\left. \overrightarrow{n}\cdot \left( -i\overrightarrow{\nabla }-%
\overrightarrow{A}\right) \right| _{\rho =R_{1}^{\ast },R_{2}^{\ast
},R_{1},R_{2}}\Psi =0\text{ .}  \eqnum{6}  \label{BC1d}
\end{equation}

Provided that Eqs.~(\ref{GL1})-(\ref{GL3}) are fulfilled, the difference
between the Gibbs free energy of the superconducting state and the normal
state is determined by the expression 
\begin{equation}
F=\frac{1}{V}\int \left[ 2\left( \overrightarrow{A}-\overrightarrow{A}%
_{0}\right) \cdot \overrightarrow{j}-\frac{\kappa _{i,o}^{2}}{\kappa _{o}^{2}%
}\left| \Psi \right| ^{4}\right] d\overrightarrow{r}\text{ ,}  \eqnum{7}
\label{F_integr}
\end{equation}
where the integral is over the total volume $V$ of the superconducting
samples and $\overrightarrow{A}_{0}=\frac{1}{2}\overrightarrow{e}_{\phi
}H_{0}\rho $ is the external vector potential in the absence of a
superconductor. The free energy is measured in $H_{c,o}^{2}V/8\pi $. The
dimensionless supercurrent $\overrightarrow{j}$ is given by $\overrightarrow{%
j}=\left( \Psi ^{\ast }\overrightarrow{\nabla }\Psi -\Psi \overrightarrow{%
\nabla }\Psi ^{\ast }\right) /2i-\left| \Psi \right| ^{2}\overrightarrow{A}$.

We restrict ourselves to samples with thickness $d<\xi $, which implies that
we are allowed to assume that the order parameter does not vary in the $z$
direction~\cite{PRL79,PRB57}. On the other hand for the vector potential the
variation in the $z$ direction is retained and for $d>\lambda $ this is very
important due to the Meissner effect~\cite{PRL79,PRB57}.

We consider sufficiently narrow rings and small disks, and therefore only
axial symmetric giant vortex states will nucleate~\cite{PRB61,PRB57}.
Consequently, the equations can be solved for a fixed value of the angular
momentum $L_{out}$ in the outer ring and $L_{in}$ in the inner
superconductor that leads to the order parameter 
\[
\Psi \left( \rho ,\phi \right) =f(\rho )e^{iL_{in,out}\phi }\text{ ,} 
\]
where $L_{in,out}=L_{out}$ in the outer ring and $L_{in,out}=L_{in}$ in the
inner superconductor. Consequently, both the vector potential and the
superconducting current density are directed along the azimuthal direction $%
\overrightarrow{e}_{\phi }$. For fixed angular momenta $L_{out}$ and $L_{in}$%
, Eqs.~(\ref{GLd1})-(\ref{GLd2}) can be reduced to the following form: 
\begin{equation}
\frac{-1}{\rho }\frac{\partial }{\partial \rho }\rho \frac{\partial f}{%
\partial \rho }+\left\langle \left( \frac{L_{in,out}}{\rho }-A\right)
^{2}\right\rangle f=f\left( \frac{\xi _{o}^{2}}{\xi _{i,o}^{2}}-\frac{\kappa
_{i,o}^{2}}{\kappa _{o}^{2}}f^{2}\right) \text{ ,}  \eqnum{8a}  \label{GLL1}
\end{equation}
\begin{equation}
-\kappa _{o}^{2}\left( \frac{\partial }{\partial \rho }\frac{1}{\rho }\frac{%
\partial \rho A}{\partial \rho }+\frac{\partial ^{2}A}{\partial z^{2}}%
\right) =\left( \frac{L_{in,out}}{\rho }-A\right) f^{2}\theta \left( \frac{%
2\left| z\right| }{d}\right) \text{ ,}  \eqnum{8b}  \label{GLL2}
\end{equation}
where $\theta (x)=1$ for $x<1$ and $\theta (x)=0$ for $x>1$, $%
\overrightarrow{A}=A\overrightarrow{e}_{\phi }$ and $\left\langle
{}\right\rangle $ indicates averaging over the disk thickness $\left\langle
f(\overrightarrow{r})\right\rangle =\frac{1}{d}\int_{-d/2}^{+d/2}f(z,\rho
)dz.$

Because the superconducting condensates of the inner and outer
superconductors are disconnected from each other they can not influence each
other directly. The coupling is entirely due to the magnetic field, or
equivalently the vector potential, as is expressed by the second term in
Eq.~(\ref{GLL1}). The total magnetic field is a sum of the applied field and
the field created by the superconducting currents of the inner ring and the
outer superconductor which is described by Eq.~(\ref{GLL2}) where $f\left(
\rho \right) =f_{in}\left( \rho \right) +f_{out}\left( \rho \right) $ and $%
f_{in}\left( \rho \right) $ $\left[ f_{out}\left( \rho \right) \right] $ is
only different from zero in the interval $R_{i}^{\ast }<\rho <R_{o}^{\ast }$ 
$\left[ R_{i}<\rho <R_{o}\right] .$

The magnetic field created by the supercurrent has a $H\sim 1/r^{3}$
dependence for large $r$. Therefore, we can restrict our calculations to a
region with radial size $R_{s}$ which we took typically 5 times the sample
size and longitudinal size $d_{s}$ typically 10 times the sample thickness .
The boundary conditions for the outer parameter can be written as 
\begin{equation}
\left. \frac{\partial f}{\partial \rho }\right| _{\rho =R_{1}^{\ast
},R_{2}^{\ast },R_{1},R_{2}}=0\text{ ,}  \eqnum{9}  \label{BCfd1a}
\end{equation}
for the ring-ring configuration, and 
\begin{equation}
\left. \frac{\partial f}{\partial \rho }\right| _{\rho =R_{2}^{\ast
},R_{1},R_{2}}=0\text{ ,\quad }\rho \left. \frac{\partial f}{\partial \rho }%
\right| _{\rho =0}=0\text{ ,}  \eqnum{10}  \label{BCfd1b}
\end{equation}
for the ring-disk configuration. The condition (\ref{BC2}) for the vector
potential taken at infinity can be transferred to the boundaries of our
simulation region 
\begin{equation}
A\left( z,\rho =R_{s}\right) =\frac{1}{2}H_{0}R_{s}\text{ ,\quad }A\left(
\left| z\right| =d_{s},\rho \right) =\frac{1}{2}H_{0}\rho \text{ .} 
\eqnum{11}  \label{BCfd2}
\end{equation}

Following the approach of Ref.~\cite{PRB57} we apply a finite difference
representation on the space grid $\rho _{n},z_{m}$ to solve Eqs.~(\ref{GLL1}%
)-(\ref{GLL2}). Since the size of our simulation region exceeds by far those
of the sample, we apply non-uniform space grids to diminish the computer
time. The space grid is taken uniform inside the sample, and we increase the
grid spacing exponentially with distance outside the sample. This allows us
to use the same number of grid points, typically 128, inside and outside the
sample. To obtain steady-state solutions for a system of two
superconductors, the following iteration procedure was used: 
\begin{eqnarray}
\eta _{f}f_{n}^{k}-\frac{2}{\rho _{n+1/2}^{2}-\rho _{n-1/2}^{2}}\left( \rho
_{n+1/2}\frac{f_{n+1}^{k}-f_{n}^{k}}{\rho _{n+1}-\rho _{n}}-\rho _{n-1/2}%
\frac{f_{n}^{k}-f_{n-1}^{k}}{\rho _{n}-\rho _{n-1}}\right) +\left\langle
\left( \frac{L_{in,out}}{\rho }-A\right) ^{2}\right\rangle _{n}f_{n}^{k} && 
\nonumber \\
-\frac{\xi _{o}^{2}}{\xi _{i,o}^{2}}f_{n}^{k}+3\frac{\kappa _{i,o}^{2}}{%
\kappa _{o}^{2}}\left( f_{n}^{k-1}\right) ^{2}f_{n}^{k}=\eta
_{f}f_{n}^{k-1}+2\frac{\kappa _{i,o}^{2}}{\kappa _{o}^{2}}\left(
f_{n}^{k-1}\right) ^{3}\text{ ,} &&  \eqnum{12a}  \label{GLfd1}
\end{eqnarray}
\begin{eqnarray}
\eta _{a}A_{m,n}^{k}-\frac{2\kappa _{o}^{2}}{\rho _{n+1/2}-\rho _{n-1/2}}%
\left( \frac{\rho _{n+1}A_{m,n+1}^{k}-\rho _{n}A_{m,n}^{k}}{\rho
_{n+1}^{2}-\rho _{n}^{2}}-\frac{\rho _{n}A_{m,n}^{k}-\rho _{n-1}A_{m,n-1}^{k}%
}{\rho _{n}^{2}-\rho _{n-1}^{2}}\right) &&  \nonumber \\
-\frac{2\kappa _{o}^{2}}{z_{m+1/2}-z_{m-1/2}}\left( \frac{%
A_{m+1,n}^{k}-A_{m,n}^{k}}{z_{m+1}-z_{m}}-\frac{A_{m,n}^{k}-A_{m-1,n}^{k}}{%
z_{m}-z_{m-1}}\right) -\left( \frac{L_{in,out}}{\rho _{n}}%
-A_{m,n}^{k}\right) \left( f_{n}^{k}\right) ^{2}=\eta _{a}A_{m,n}^{k-1}\text{
,} &&  \eqnum{12b}
\end{eqnarray}
where $A_{m,n}=A(z_{m},\rho _{n})$, $f_{n}=f\left( \rho _{n}\right) $, $\rho
_{n+1/2}=\left( \rho _{n+1}+\rho _{n}\right) /2$ and $z_{m+1/2}=\left(
z_{m+1}+z_{m}\right) /2$. The upper index $k$ denotes the iteration step. To
speed up the convergency we introduced the iteration parameters $\eta _{f}$
and $\eta _{a}$, and we expanded the nonlinear term $\left( f_{n}^{k}\right)
^{3}=\left( f_{n}^{k-1}\right) ^{3}+3\left( f_{n}^{k-1}\right) ^{2}\left(
f_{n}^{k}-f_{n}^{k-1}\right) .$

\section{Ring-disk configuration}

First, we consider a superconducting ring with a superconducting disk in the
center. We investigate the influence of the disk on the properties of the
ring. As an example, we take a ring with inner radius $R_{i}=1.5\xi $ and
outer radius $R_{o}=2.0\xi $ and a disk in the center with radius $%
R_{o}^{\ast }=1.0\xi $. Both superconductors have the same thickness, $%
d=0.15\xi $, and the Ginzburg-Landau parameter was taken to be $\kappa =0.28$%
, which is a typical value for mesoscopic Al superconductors~\cite
{Geim1,PRL79}.

\begin{figure}[tbp]
\begin{center}
\epsfig{file=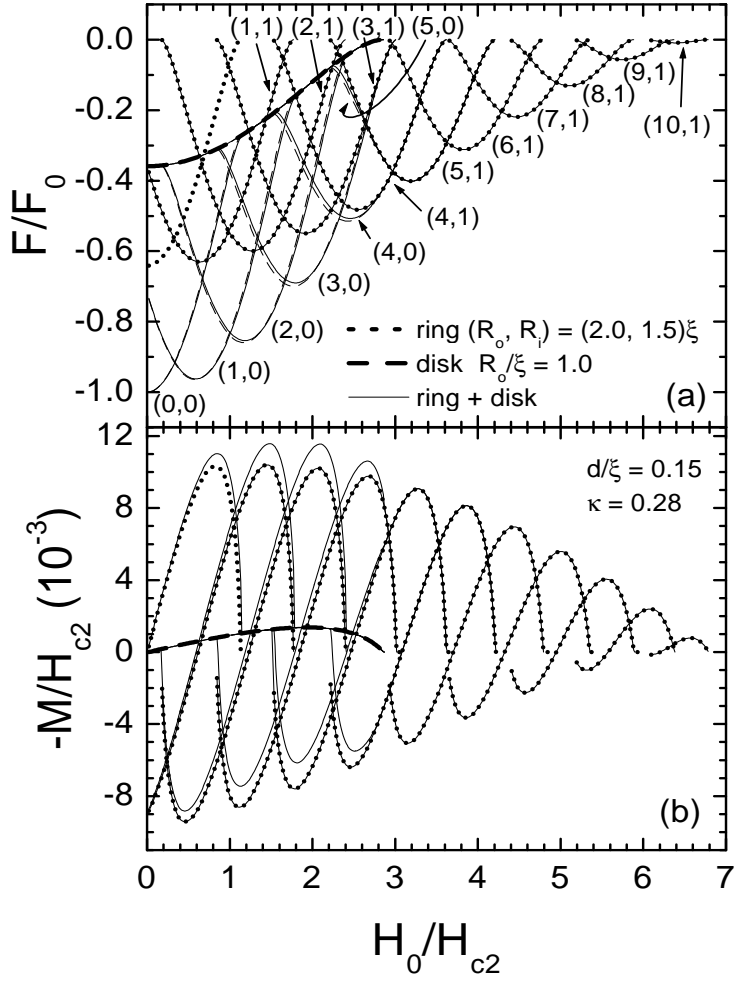, width=110mm, height=150mm, clip=}
\end{center}
\caption{(a) The free energy and (b) the expelled field as a function of the
applied field for a ring with inner radius $R_{i}=1.5\protect\xi $ and outer
radius $R_{o}=2.0\protect\xi $ (thick dotted curves) and a disk in the
center with radius $R_{o}^{\ast }=1.0\protect\xi $ (thick dashed curves) and
for the coupled ring-dot configuration (thin solid curves). All
superconductors have the same thickness, $d=0.15\protect\xi $, and
$\protect%
\kappa =0.28$. The thin dashed curves give the sum of the free energies of
the single disk and the single ring.}
\label{emagrd}
\end{figure}

Fig.~\ref{emagrd}(a) shows the free energy for the considered system as a
function of the applied magnetic field. First we consider the uncoupled
system and calculated the free energy for the disk in the center (thick
dashed curve) and the free energy for the different giant vortex states in
the outer ring (thick dotted curve). The results for single rings and disks
were exhaustively described in previous papers~\cite{PRB57,PRB61}. Notice
that, in this paper, the free energy is expressed in units of $%
F_{0}=H_{c}^{2}V/8\pi $, where $V$ is the sum of the disk and the ring
volume. This is the reason why the free energy of the disk and the ring are
not equal to $-F_{0}$ at zero magnetic field as it was in Refs.~\cite
{PRB57,PRB61} The size of the disk is such that only the Meissner state,
i.e. the $L_{in}=0$ state, can nucleate. At applied magnetic fields $%
H_{0}/H_{c2}\gtrsim 2.9$ the disk is in the normal state, which results in $%
F=0$. In the single ring, on the other hand, different giant vortex states
with vorticity $L_{out}=0$ up to $L_{out}=10$ can nucleate before the ring
becomes normal at $H_{0}/H_{c2}\approx 6.8$. Next, we introduced the
magnetic coupling between the disk and the ring and the results are given by
the solid curves in Fig.~\ref{emagrd}(a). The different axial symmetric
states are determined by the vorticity of the disk $L_{in}$ and the total
vorticity $L_{out}$, which is equal to the vorticity of the ring. Therefore,
we characterize the states by $\left( L_{out},L_{in}\right) $. For the
considered configuration, we find states with $L_{in}=0$ and $L_{out}=0$ up
to $L_{out}=5.$ We also find states with $L_{in}=1$ and $L_{out}=0$ up to $%
L_{out}=10$, which equal the giant vortex states of the single ring, because
the disk is now in the normal state. Notice further that we could also write 
$\left( 6,0\right) $ instead of $(6,1)$ because for the applied magnetic
fields where the $L_{out}=6$ state in the ring exists the disk is normal
even for $L_{in}=0.$ We have chosen to write $L_{in}=1$ because this
expresses more clearly that there is flux going through the disk. If both
the disk and the ring are superconducting, the free energy of the total
sample is different from the sum of the free energies of the single disk and
the single ring.

\begin{figure}[tbp]
\begin{center}
\epsfig{file=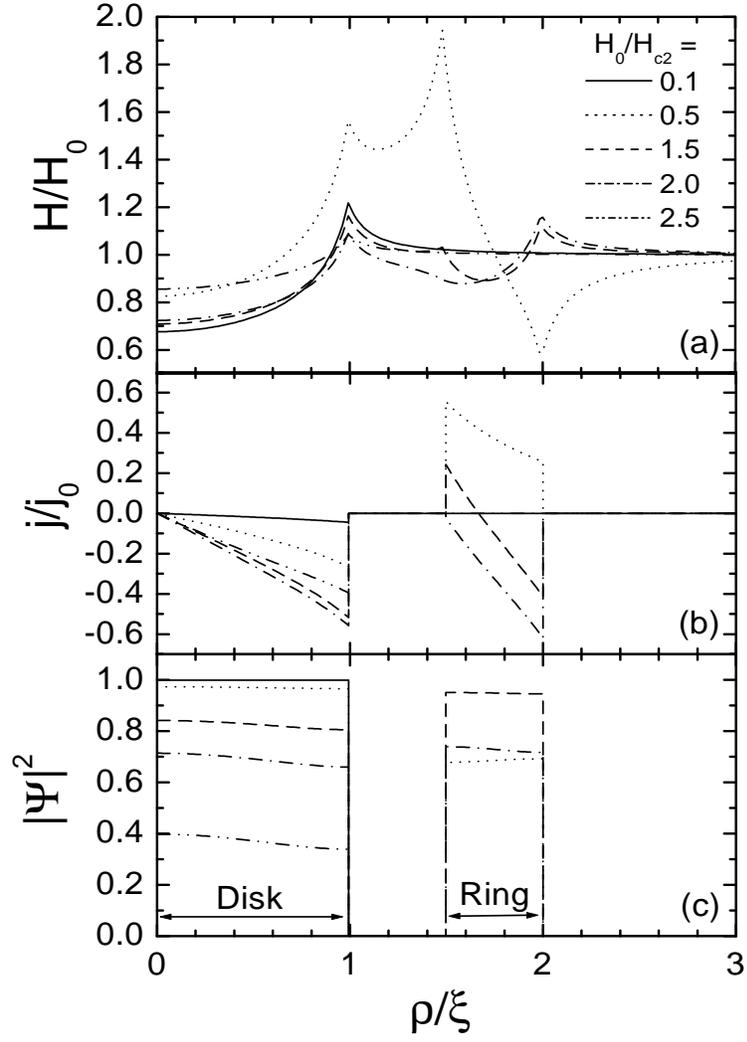, width=110mm, height=150mm, clip=}
\end{center}
\caption{(a) The magnetic field distribution, (b) the current density and
(c) the Cooper-pair density as a function of the radial position for the $%
\left( L_{out},L_{in}\right) =\left( 2,0\right) $ state of the ring-dot
configuration of Fig.~\ref{emagrd} at $H_{0}/H_{c2}=0.1$, $0.5$, $1.5$,
$2.0$%
, and $2.5$.}
\label{curretc}
\end{figure}

To investigate these new states in more detail we consider as an example the 
$\left( 2,0\right) $ state. Figs.~\ref{curretc}(a-c)\ show the magnetic
field distribution, the current density and the Cooper-pair density,
respectively, as a function of the radial position for five different
applied magnetic fields, i.e., $H_{0}/H_{c2}=0.1$, $0.5$, $1.5$, $2.0$, and $%
2.5.$ Near $H_{0}/H_{c2}=0$ the $\left( 2,0\right) $ state equals the $%
L_{in}=0$ state of the disk and the ring is in the normal state. The reason
is that the applied field is so low that a lot of magnetic flux has to be
attracted to create a state with $L_{out}=2$ in the outer ring. Therefore, a
very high superconducting current has to flow through the outer ring which
leads to the destruction of superconductivity in this ring. The solid curves
in Figs.~\ref{curretc}(a-c) show that the Cooper-pair density and the
current density are indeed zero in the ring. The magnetic field distribution
shows the flux expulsion from the disk. Inside the disk the field decreases
and at the edge there is a peak which illustrates a higher concentration of
field because of the demagnetization effects. With increasing external field
less flux has to be attracted and the current in the outer ring decreases.
At $H_{0}/H_{c2}\approx 0.17$ superconductivity is restored in the external
ring (see the dotted curves in Fig.~\ref{curretc}(c)). The dotted curves in
Fig.~\ref{curretc}(b) show that the current in the outer ring flows in the
opposite direction than the current in the disk. The superconducting
currents in the disk expel the flux, while the currents in the ring are
attracting flux, which is compressed in the region between the disk and the
ring (see the dotted curve in Fig.~\ref{curretc}(a)). The free energy
becomes now more negative as compared to the free energy of the single disk
(see Fig.~\ref{emagrd}(a)). Increasing the magnetic field further leads to
less attraction of flux and, hence, to a higher Cooper-pair density in the
ring and a more negative free energy. When the external flux becomes
comparable with the flux needed for the $L_{out}=2$ state, the outer part of
the ring expells the flux to the outside, while the inner part of the ring
still expells flux to the hole region. Therefore, the superconducting
current changes sign in the ring region (see the dashed curve in Fig.~\ref
{curretc}(b)). Since the flux is expelled in both directions, the dashed
curve in Fig.~\ref{curretc}(a) shows a positive peak at both ring
boundaries. Further increasing the external field leads to external fluxes
larger than the flux needed for the $L_{out}=2$ state and, hence, the ring
has to expel flux in order to keep vorticity $L_{out}=2$. As a consequence,
the current in the ring has to flow in the same direction as the current in
the disk, which is shown by the dash-dotted curve in Fig.~\ref{curretc}(b).
Because of the expulsion, the field between the two superconductors is lower
than the external field (see the dash-dotted curve in Fig.~\ref{curretc}%
(a)). If we further increase the magnetic field, the superconducting
currents in the outer ring have to increase in order to expel more flux and
consequently the Cooper-pair density decreases in the outer ring. At $%
H_{0}/H_{c2}\approx 2.4$ the supercurrent becomes too high and the ring
becomes normal again (see the dash-dot-dotted curves in Figs.~\ref{curretc}%
(a-c)). At this field, the free energy equals the free energy of the single
disk.

The above discussion shows clearly the interplay between the superconducting
state of the disk and the ring. Next, we investigate the interaction between
the disk and the ring. Therefore, we added the sum of the free energies of
the single disk and the single ring (thin dashed curves in Fig.~\ref{emagrd}%
(a)) and compare this result with the result of the ring-disk configuration
(solid curves in Fig.~\ref{emagrd}(a)). Notice that there is a small
difference between the two set of curves, which is due to the coupling
between the two superconductors. Below we will show that this difference
becomes more pronounced for thicker samples.

Now, we will determine the attraction or expulsion of the magnetic field by
the coupled superconducting system. Fig.~\ref{emagrd}(b) shows the magnetic
field expelled from the sample, $-M$, as a function of the applied magnetic
field: 
\[
M=\frac{\left\langle H\right\rangle -H_{0}}{4\pi }\text{ ,} 
\]
where $\left\langle H\right\rangle $ is the magnetic field averaged over the
area $\rho <R_{o}$, i.e. the outer radius of the ring and $H_{0}$ is the
applied field. The thick dashed and thick dotted curves are the results for
the single disk and the single ring and the solid curves for the total
ring-disk system. By putting a disk in the center of the ring, more field is
expelled and less field is attracted. Of course, for $H_{0}/H_{c2}\gtrsim
2.9 $ the disk is in the normal state and we recover the result for the
single ring case.

\section{Ring-ring configuration}

In this section we replace the disk in the center by a second ring and the
influence of this inner ring on the outer ring will be investigated. As an
example for this ring-ring configuration, we consider a superconducting ring
with inner radius $R_{i}=1.5\xi $ and outer radius $R_{o}=2.0\xi $ and a
second ring in the center with inner radius $R_{i}^{\ast }=0.6\xi $ and
outer radius $R_{o}^{\ast }=1.1\xi $. Both rings have the same thickness, $%
d=0.15\xi $ and the same Ginzburg-Landau parameter $\kappa =0.28$.

\begin{figure}[tbp]
\begin{center}
\epsfig{file=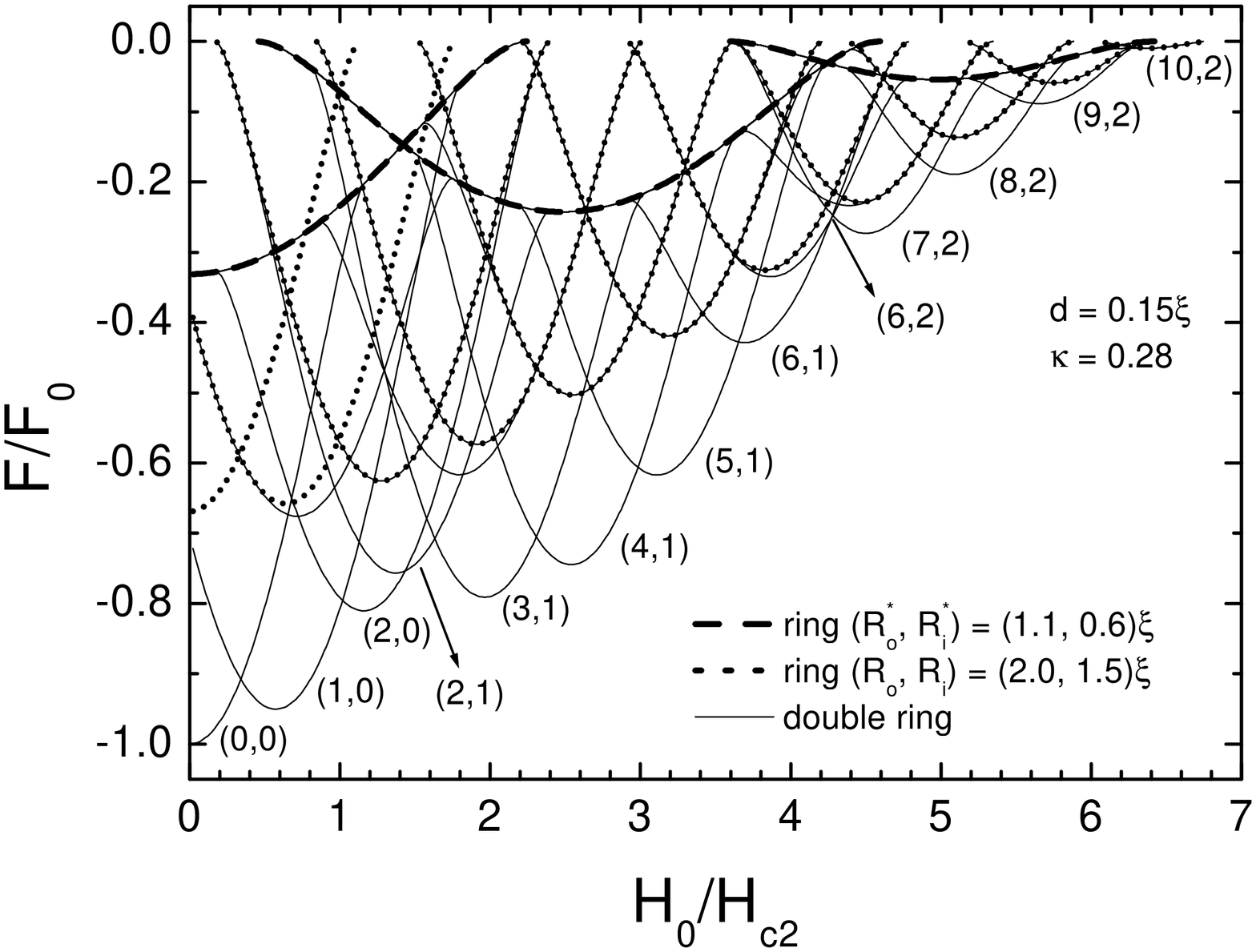, width=150mm, height=110mm, clip=}
\end{center}
\caption{The free energy as a function of the applied magnetic field for a
superconducting ring with radii $R_{i}=1.5\protect\xi $ and $R_{o}=2.0%
\protect\xi $ (dotted curves), a ring with $R_{i}^{\ast }=0.6\protect\xi $
and $R_{o}^{\ast }=1.1\protect\xi $ (dashed curves) and the double ring
configuration (solid curves). Both rings have the same thickness, $d=0.15%
\protect\xi $, and $\protect\kappa =0.28$. The indices indicate the ground
state vorticities $\left( L_{out},L_{in}\right) $.}
\label{emagrr}
\end{figure}

Fig.~\ref{emagrr} shows the free energy as a function of the applied
magnetic field for the small single ring by thick dashed curves, for the
larger single ring by thick dotted curves and for the coupled ring-ring
situation by the thin solid curves, where the interaction between the two
rings is taken into account. In the inner ring superconducting states can
nucleate with vorticity $L_{in}=0$, $1$ and $2$ and the
superconducting/normal transition field is at $H_{0}/H_{c2}\approx 6.4$. In
the outer ring states with vorticity $L_{out}=0$ up to $L_{out}=10$ exist
and superconductivity is destroyed at $H_{0}/H_{c2}\approx 6.75$. The
superconducting states nucleating in the double ring system can be
characterized again by the indices $\left( L_{out},L_{in}\right) $. For $%
L_{in}=0$, superconducting states can nucleate with $L_{out}=0$ up to $4$,
for $L_{in}=1$ with $L_{out}=1$ up to $L_{out}=8,$ and for $L_{in}=2$ with $%
L_{out}=5$ up to $10$. For $L_{in}\geqslant 3$ the states equal the states
of the single outer ring because the inner ring will be normal.

The indices in the figure correspond to the ground state of the coupled ring
system. For the numerical example studied in the previous section the number
and the position of the ground state transitions are the same as for the
outer ring. In the present two ring system this is no longer the case and
the number of transitions in the coupled system are larger than for the
single outer ring case. The inner ring induces extra transitions in the
coupled system each time when the vorticity of the inner ring changes with
one unit. The first extra transition is the transition from $(2,0)$ to $%
(2,1) $ and the second one is the transition from $(6,1)$ to $(6,2)$. The $%
(2,1)$ state is the ground state in the magnetic field region $1.43\lesssim
H_{0}/H_{c2}\lesssim 1.63$ and the $(6,2)$ state in the region $4.27\lesssim
H_{0}/H_{c2}\lesssim 4.28$. Hence, by putting a ring in the center of the
larger ring, the ground state shows extra transitions. This result
corresponds to the experimental result of Morelle {\it et al}~\cite{Morelle}%
, who saw modifications of the $T_{c}(H)$ oscillations of the outer ring in
the Fourier spectrum of the $T_{c}(H)$ line due to the coupling between the
outer and the inner ring.

\begin{figure}[tbp]
\begin{center}
\epsfig{file=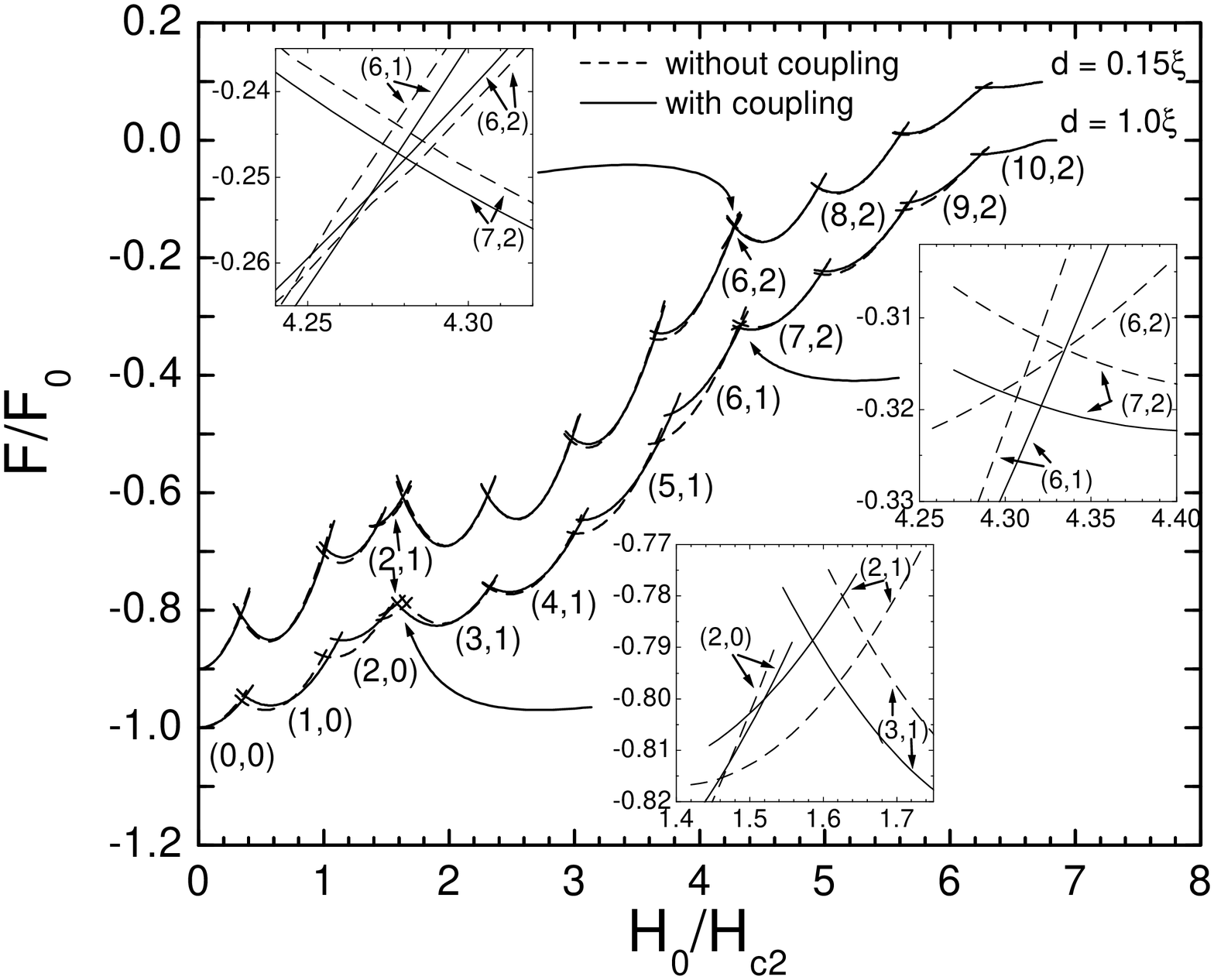, width=150mm, height=110mm, clip=}
\end{center}
\caption{The ground state free energy of the double ring configuration of
Fig.~\ref{emagrr} (solid curves) and the sum of the free energies of the two
rings (dashed curves) for $d=0.15\protect\xi $ (upper curves) and for
$d=1.0%
\protect\xi $ (lower curves). The upper curve is shifted by $+0.1$ for
clarity. The insets show some of the crossings in more detail.}
\label{emaggrrr}
\end{figure}

Next, we focus further on the interaction between the inner ring and the
outer ring. Therefore, we plot in Fig.~\ref{emaggrrr} the ground state free
energy of the coupled rings (solid curves) and the sum of the free energies
of the two single rings (dashed curves) for the previous configuration
(upper curves which are shifted by $+0.1$), i.e. $d=0.15\xi $, and for a
thicker sample with $d=1.0\xi $ (lower curves). For $d=0.15\xi $ the
difference is most pronounced for the $\left( 2,0\right) $, the $\left(
5,1\right) $ and the $\left( 6,1\right) $ state. Both the value of the free
energy and the transition magnetic fields are influenced by the interaction
between the two rings. The left inset shows the $\left( 6,1\right)
\rightarrow \left( 6,2\right) \rightarrow \left( 7,2\right) $ transition in
more detail. Notice that the interaction significantly decreases the
magnetic field region where the $\left( 6,2\right) $ state is the ground
state. For $d=1.0\xi $ the demagnetization effects become more important and
therefore the interaction between the two rings gains importance. This
results in a larger difference between the dashed and solid curves. The
value of the free energy and the transition fields are changing considerably
by the fact that both rings are influencing each other. The two lower insets
show the $\left( 2,0\right) \rightarrow \left( 2,1\right) \rightarrow \left(
3,1\right) $ and the $\left( 6,1\right) \rightarrow \left( 7,2\right) $\
transitions in more detail. The magnetic field region where the ground state
is given by the $\left( 2,1\right) $ state decreases due to the interaction.
For the $\left( 6,1\right) \rightarrow \left( 7,2\right) $ transition
coupling between the two rings leads to the interesting result that the $%
\left( 6,2\right) $ state is no longer a ground state.

\begin{figure}[tbp]
\begin{center}
\epsfig{file=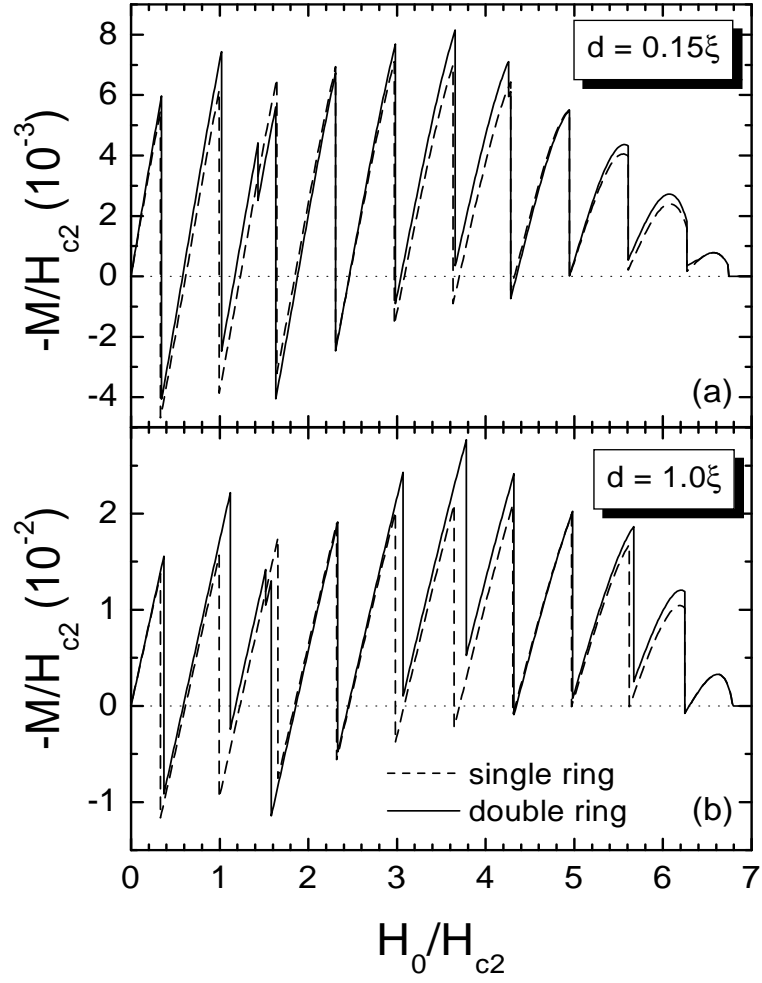, width=110mm, height=150mm, clip=}
\end{center}
\caption{The magnetic field expelled from the sample, $-M$, as a function of
the applied magnetic field for the single outer ring (dashed curve) and for
the double ring configuration (solid curve). The sample is the same as in
Fig.~\ref{emagrr} with thickness $d=0.15\protect\xi $ (a) and
$d=1.0\protect%
\xi $ (b).}
\label{magnrr}
\end{figure}

Figs.~\ref{magnrr}(a,b) shows the magnetic field expelled from the sample, $%
-M$, as a function of the applied magnetic field for the single outer ring
(dashed curve) and for the double ring configuration (solid curve) for
thickness $d=0.15\xi $ and $d=1.0\xi $ respectively. For $d=0.15\xi $, the
two extra transitions, resulting from the influence of the inner ring, are
clearly visible by the jumps at $H_{0}/H_{c2}=1.43$ and $4.28$. Notice that,
depending on the direction of the current a single ring expels or attracts
field at a given applied magnetic field and vorticity. Expulsion
(attraction) leads to a lower (higher) magnetic field density in the center
of the ring and to a higher (lower) density near the outside. Therefore,
depending on the applied field, the field expulsion or attraction in the
coupled ring configuration can either increase if the two currents are in
the same direction, or decrease if the directions are opposite. This can be
clearly seen from Figs.~\ref{magnrr}(a,b) where the field expulsion or
attraction becomes more pronounced by putting the inner ring in the center
of the outer ring. For $d=1.0\xi $ one extra transition results from the
influence of the inner ring, i.e. at $H_{0}/H_{c2}=1.58$. The difference
between the transition fields of the single ring and the coupled ring system
becomes larger (see also the lower curve in Fig.~\ref{emaggrrr}) and the
difference in the expulsion becomes more pronounced with increasing the
sample thickness, which indicates again that the interaction between the two
rings increases with increasing $d$.

\begin{figure}[tbp]
\begin{center}
\epsfig{file=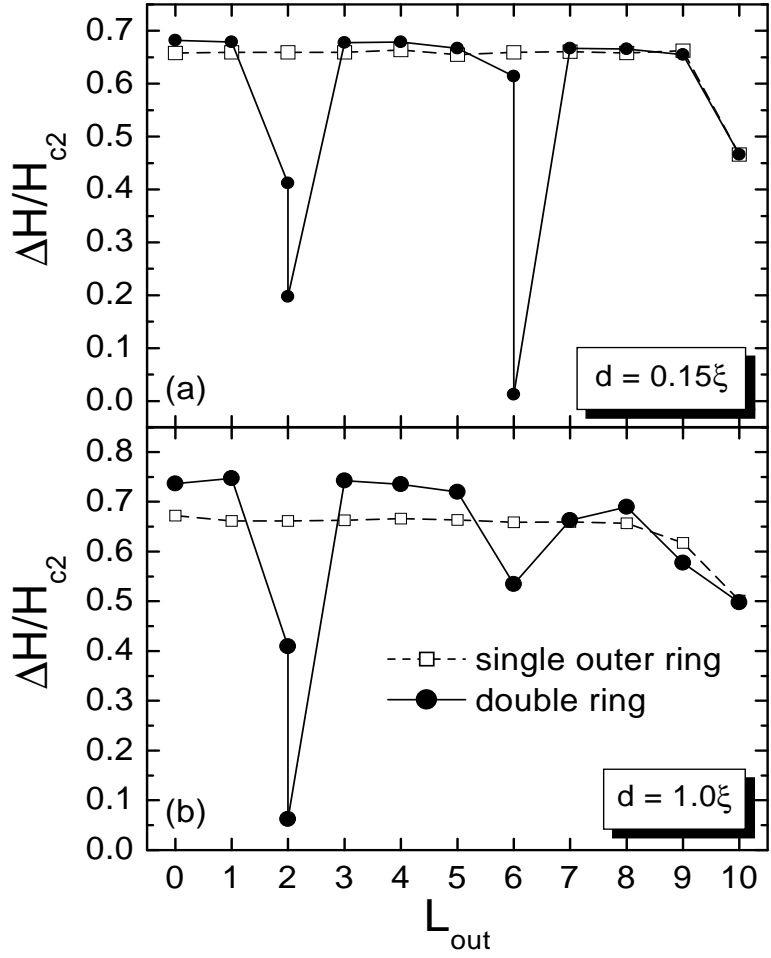, width=110mm, height=150mm, clip=}
\end{center}
\caption{The magnetic field range, $\Delta H_{0}$, over which the $\left(
L_{out},L_{in}\right) $ state is the ground state, as a function of
$L_{out}$
for the single outer ring (the dashed curve and open squares) and for the
double ring system (the solid curve and closed circles). The sample is the
same as in Fig.~\ref{emagrr} with thickness $d=0.15\protect\xi $ (a) and $%
d=1.0\protect\xi $ (b).}
\label{deltahrr}
\end{figure}

In Figs.~\ref{deltahrr}(a,b) the magnetic field range, $\Delta H_{0}$, over
which the $\left( L_{out},L_{in}\right) $ state is the ground state, is
plotted as a function of $L_{out}$ for thickness $d=0.15\xi $ and $d=1.0\xi $
respectively.\ This magnetic field range corresponds to the distance between
two consecutive jumps in the expelled field (see Fig.~\ref{magnrr}). The
results for the single outer ring are given by the open squares and for the
double ring system by the closed circles. The curves are guides to the eye.
For $d=0.15\xi $ the extra transitions are clearly visible at $L_{out}=2$
and $6$ and for $d=1.0\xi $ at $L_{out}=2$ where for the same $L_{out}$ two
jumps occur due to a transition of the inner ring. Also for the other
vorticities $L_{out}$, there is a difference between $\Delta H_{0}$ for the
single ring and the double ring. The reason is that the ground state
transition fields are influenced by the interaction between the two rings.
This was also visible in Fig.~\ref{emaggrrr}. If the free energy of the
double ring was just the sum of the free energies of the two single rings, $%
\Delta H_{0}$ would be the same for the single outer ring and the double
ring, except for $L_{out}=2$ and $6$, where extra transitions occur because $%
L_{in}$ changes with one unit. Notice further that the difference between
the results for the single outer ring and the double ring enhances with
increasing sample thickness.

\begin{figure}[tbp]
\begin{center}
\epsfig{file=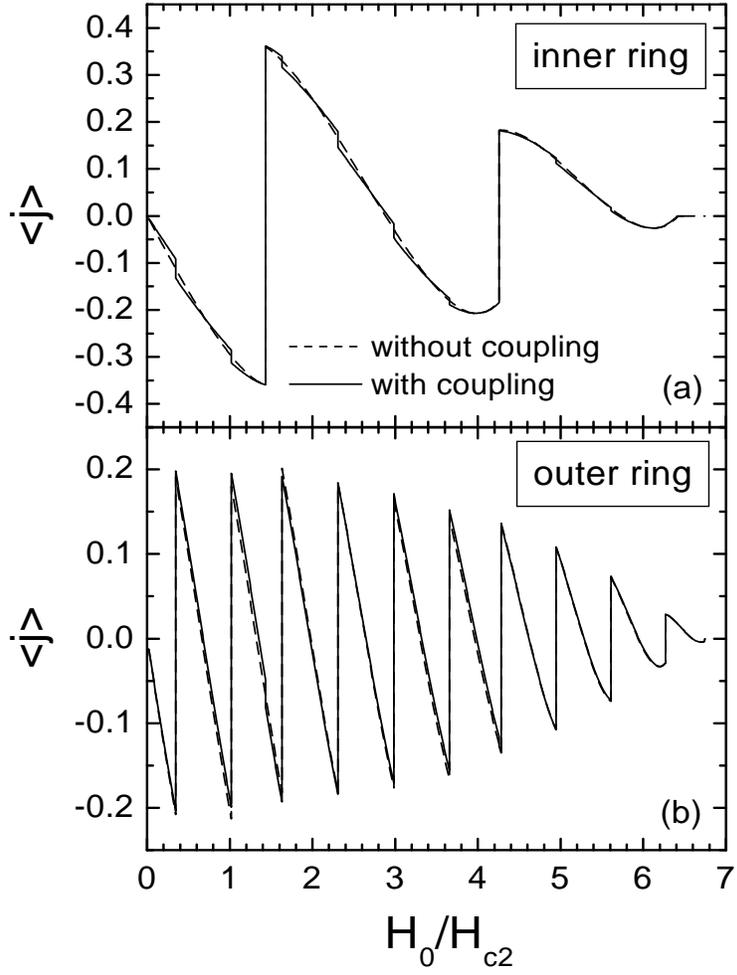, width=110mm, height=150mm, clip=}
\end{center}
\caption{The averaged current density for the ground state in the inner ring
(a) and the outer ring (b) as a function of the applied magnetic field for
the same double ring configuration as in Fig.~\ref{emagrr}. The results for
the single rings are given by dashed curves, these for the double ring
configuration by solid curves.}
\label{currentrr}
\end{figure}

Next, we investigate the effect of the interaction between the two rings on
the superconducting current density in the two rings for $d=0.15\xi $. Figs.~%
\ref{currentrr}(a,b) show the averaged current density for the ground state
in the inner ring and the outer ring, respectively, as a function of the
applied magnetic field. The results for the single ring are given by dashed
curves, these for the double ring configuration by solid curves. First, we
describe what happens if there is no interaction between the two rings. In
this case we can consider them as two single rings. At low fields, the
ground state of a single ring is given by the $L=0$ state or Meissner state
and the ring expels the field to the outside of the sample. With increasing
external field, more flux has to be expelled from the ring which leads to a
higher current density. After the first transition the ground state is given
by the $L=1$ state and initially flux will be trapped in the ring and the
flux going through the ring is larger than the flux of the external field.
To compress this extra magnetic field, the superconducting current in the
ring has to flow in the opposite direction. At the transition, the current
shows a jump from a negative to a positive value, i.e. from expulsion to
compression. With increasing external field, less flux has to be compressed
to achieve vorticity one and the current density in the outer ring
decreases. Further increasing the field, the external flux becomes larger
than the flux needed for $L=1$ and flux has to be expelled. Therefore, the
current in the ring changes sign. Without interaction between the two rings,
the current density in one ring exhibits only jumps when the vorticity of
the ground state of this ring changes (see dashed curves in Figs.~\ref
{currentrr}(a,b)).

In the coupled two rings situation $\left\langle j\right\rangle $ shows
small jumps on top of the previously described expulsion $\rightarrow $
compression jumps. At low fields, the ground state is given by the $\left(
0,0\right) $ state or Meissner state. Both rings expel the field to the
outside of the sample, which means that the current flows in the same
direction in each ring. Since some flux is already expelled by the outer
ring, the inner ring has to expel less and therefore the current is less
negative. After the first transition the groundstate changes into the $%
\left( 1,0\right) $ state. Now, the outer ring compresses the field to
achieve vorticity one, and, as a consequence, the field in the hole of the
outer ring is larger than the external field. This means that the inner ring
has to expel more field and the current density jumps to a value more
negative than its value without interaction. The other transitions can be
explained analogously. From Figs.~\ref{currentrr}(a,b) it is clear that the
two rings are influencing each other and that the interaction between the
two rings results in extra jumps in the current density in one ring when the
vorticity of the other ring changes. These jumps are smaller than the jumps
when the vorticity of the considered ring increases, but they are not
negligible.

\begin{figure}[tbp]
\begin{center}
\epsfig{file=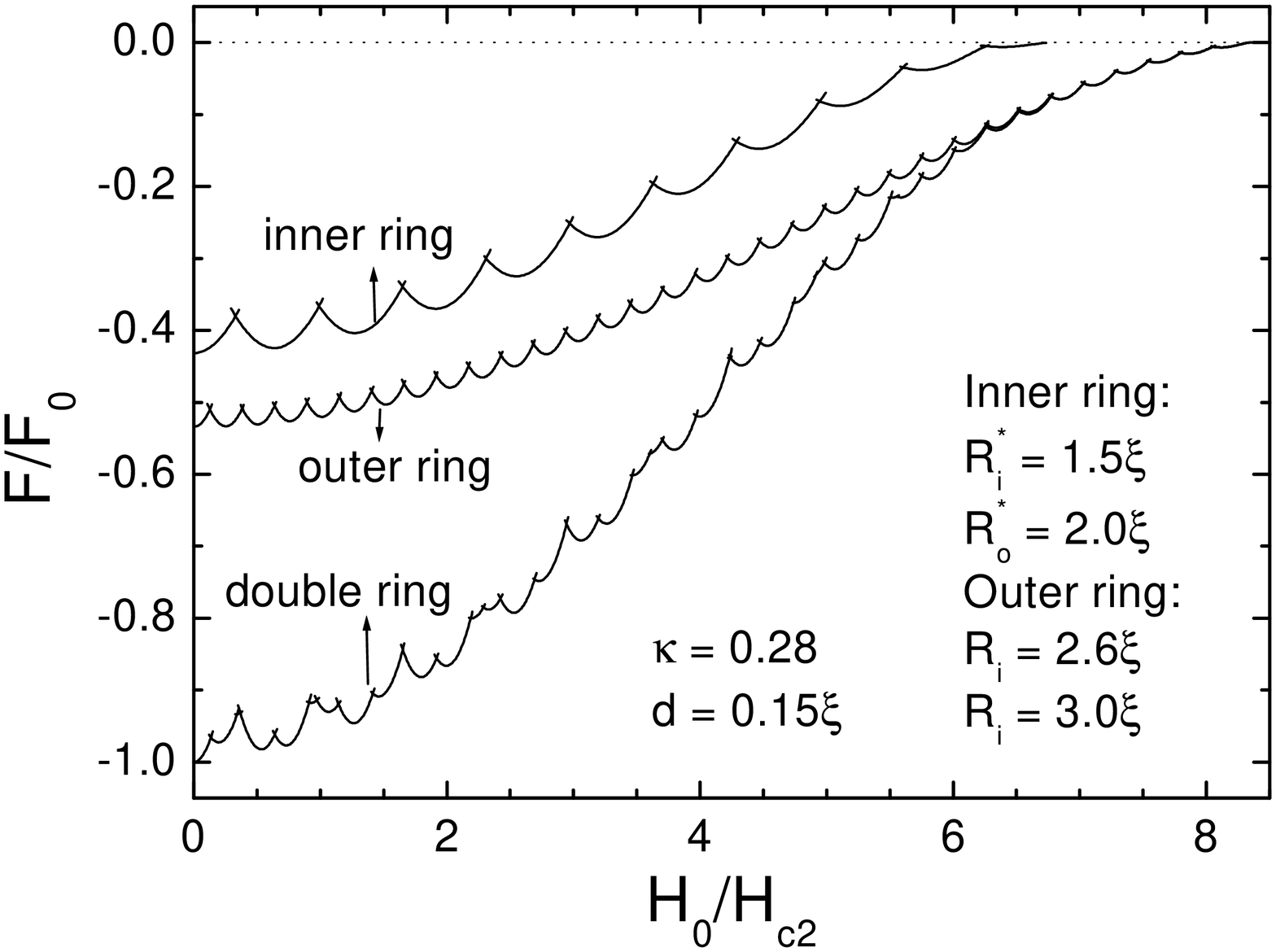, width=150mm, height=110mm, clip=}
\end{center}
\caption{The ground state free energy for a single inner ring with radii $%
R_{o}^{\ast }=2.0\protect\xi $ and $R_{i}^{\ast }=1.5\protect\xi $, for a
single outer ring with radii $R_{o}=3.0\protect\xi $ and $R_{i}=2.6\protect%
\xi $ and for the double ring configuration, i.e. the combination of these
two rings. The sample thickness is $d=0.15\protect\xi $ and the
Ginzburg-Landau parameter $\protect\kappa =0.28$. }
\label{engr}
\end{figure}

Up to now, we considered rather small samples. For the single ring it is
known that by increasing the sample size (i) more $L$ states are possible
and (ii) the magnetic field range over which the state with vorticity $L$ is
the ground state decreases~\cite{PRB61}. Therefore, for a larger radius of
the double ring configuration, we expect many more ground state transitions.
Fig.~\ref{engr} shows the ground state free energy for a single inner ring
with radii $R_{o}^{\ast }=2.0\xi $ and $R_{i}^{\ast }=1.5\xi $, for a single
outer ring with radii $R_{o}=3.0\xi $ and $R_{i}=2.6\xi $ and for the
coupled ring-ring configuration. The sample thickness is $d=0.15\xi $ and
the Ginzburg-Landau parameter $\kappa =0.28$. For the single inner ring, the
ground state changes from vorticity $L_{in}=0$ up to $L_{in}=10$ and the
superconducting/normal transition field is at $H_{0}/H_{c2}=6.73.$ For the
single outer ring the ground state changes from vorticity $L_{out}=0$ up to $%
L_{out}=32$ and superconductivity is destroyed at $H_{0}/H_{c2}=8.40$. By
comparing the free energy of the double ring with the one of the outer ring,
we notice that there are many more ground state transitions as a consequence
of the transitions in the inner ring. For the single ring the minimum in the
free energy of the $L+1$ state is always less negative than the one of the $%
L $ state. Due to the interaction between the two rings, this is no longer
always the case for the double ring configuration. At $H_{0}/H_{c2}>6.73$
the free energy of the double ring configuration equals the one of the outer
ring since the inner ring is in the normal state.

\begin{figure}[tbp]
\begin{center}
\epsfig{file=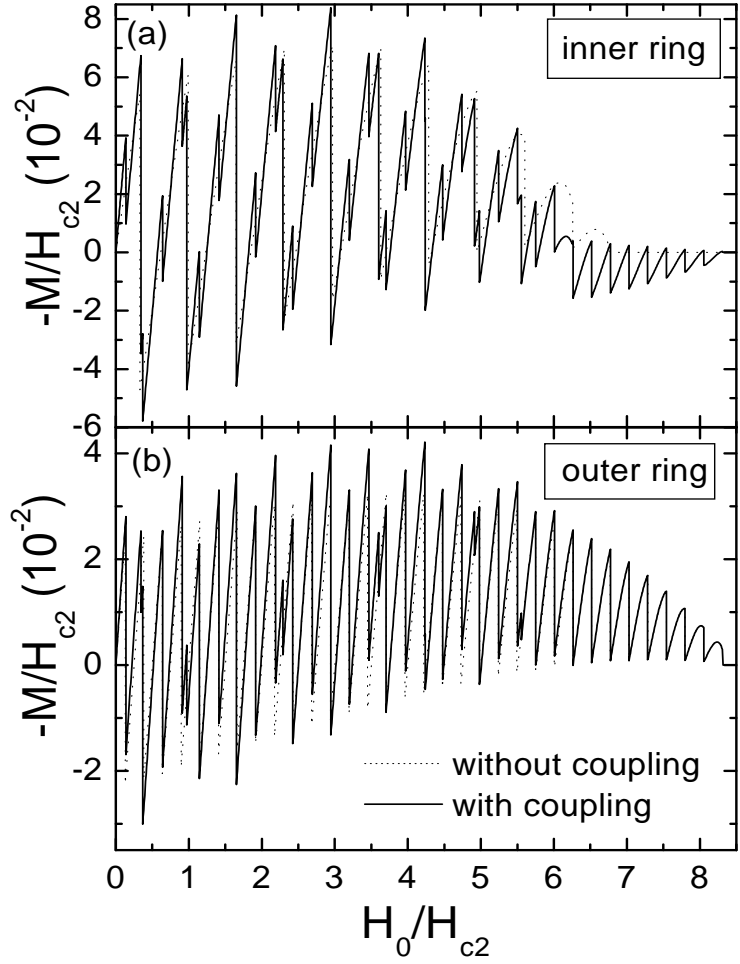, width=110mm, height=150mm, clip=}
\end{center}
\caption{The field expelled from the region $\protect\rho <R_{o}^{\ast }$ of
the inner ring with radii $R_{o}^{\ast }=2.0\protect\xi $ and $R_{i}^{\ast
}=1.5\protect\xi $ (a), and the region $\protect\rho <R_{o}$ of the outer
ring with radii $R_{o}=3.0\protect\xi $ and $R_{i}=2.6\protect\xi $ (b). The
results for single rings are given by the dotted curves and for the double
ring configuration by the solid curves. $(d=0.15\protect\xi $ and $\protect%
\kappa =0.28)$.}
\label{magngr}
\end{figure}

Figs.~\ref{magngr}(a,b) show the field expelled from the region $\rho
<R_{o}^{\ast }$ of the inner ring with radii $R_{o}^{\ast }=2.0\xi $ and $%
R_{i}^{\ast }=1.5\xi $ and the region $\rho <R_{o}$ of the outer ring with
radii $R_{o}=3.0\xi $ and $R_{i}=2.6\xi ,$ respectively. The results for
single rings are given by the dotted curves and for the double ring
configuration by the solid curves. At low fields, the single inner ring is
in the Meissner state and expels the magnetic field, i.e. $-M>0$. With
increasing external field, more field is expelled. At $H_{0}/H_{c2}\approx
0.33$ the ground state changes from $L_{in}=0$ to $L_{in}=1$ and flux has to
be compressed into the hole to achieve vorticity one. Therefore, the
magnetic field inside the hole will be larger than the external one and $-M$
jumps to negative values, i.e. field compression. With increasing field,
less flux has to be attracted and $-M$ becomes less negative. Further
increasing the external flux becomes larger than the one needed for $%
L_{in}=1 $ and, therefore, the field has to be expelled again. This means
that $-M$ becomes positive. Further increasing the field, more flux has to
be expelled and $-M$ becomes more positive. At $H_{0}/H_{c2}\approx 0.99$
the vorticity changes from $L_{in}=1$ to $L_{in}=2$, which means that $-M$
jumps to negative values, and so forth. At $H_{0}/H_{c2}=6.73$
superconductivity is destroyed and the field becomes equal to the external
one, and as a consequence, $-M=0$. The description for the single outer ring
is completely analogous.

Placing a larger ring around the inner ring influences the expelled field
drastically (see the solid curves in Fig.~\ref{magngr}(a)). Due to the
expulsion of the outer ring at low fields, the magnetic field inside the
hole of this ring will be smaller than the external one. Now, the expulsion
by the inner ring results in a smaller local field than for the case of the
single ring and, as a consequence, the expelled field increases at low
fields. At $H_{0}/H_{c2}\approx 0.13$ the ground state of the outer ring
changes from vorticity $L_{out}=0$ to $L_{out}=1$, which means that suddenly
the outer ring has to attract flux to achieve vorticity one. Therefore, the
field inside the hole of the outer superconductor becomes larger than the
external one and the expulsion by the inner ring will be less pronounced
than for the case of the single ring. As a consequence, $-M$ jumps from a
value above the one for the single ring case to a value below this value.
This interplay between the two rings leads to a higher expulsion
(attraction) from the region $\rho <R_{o}^{\ast }$ when the outer ring
expels (attracts) flux and to a lower expulsion when the outer ring attracts
(expels) flux. For the outer ring an analogous explanation can be given (see
the solid curves in Fig.~\ref{magngr}(b)).

\section{Two coupled rings of different materials}

Until now, we considered always two superconductors made of the same
material. This means that both superconductors have the same coherence
length, penetration depth and critical temperature, i.e. $\xi _{i}=\xi _{o}$%
, $\lambda _{i}=\lambda _{o}$, and $T_{c,i}=T_{c,o}$. Since both rings have
the same width and the radius of the inner ring is smaller than the one of
the outer ring, the inner ring becomes normal at a smaller field than the
outer ring. As a consequence, no effect of the magnetic coupling can be
observed in the $H-T$ phase diagram. To circumvent this problem the $T_{c}$
of the outer ring was artificially lowered in the experiment of Ref.~\cite
{Morelle} by applying a sufficiently large external current through the
outer ring. An alternative approach will be followed in the present section
where we take the inner ring of a different material such that it has a
higher critical temperature than the outer ring, and also a different
coherence length and penetration depth, which leads to a different
Ginzburg-Landau parameter.

\begin{figure}[tbp]
\begin{center}
\epsfig{file=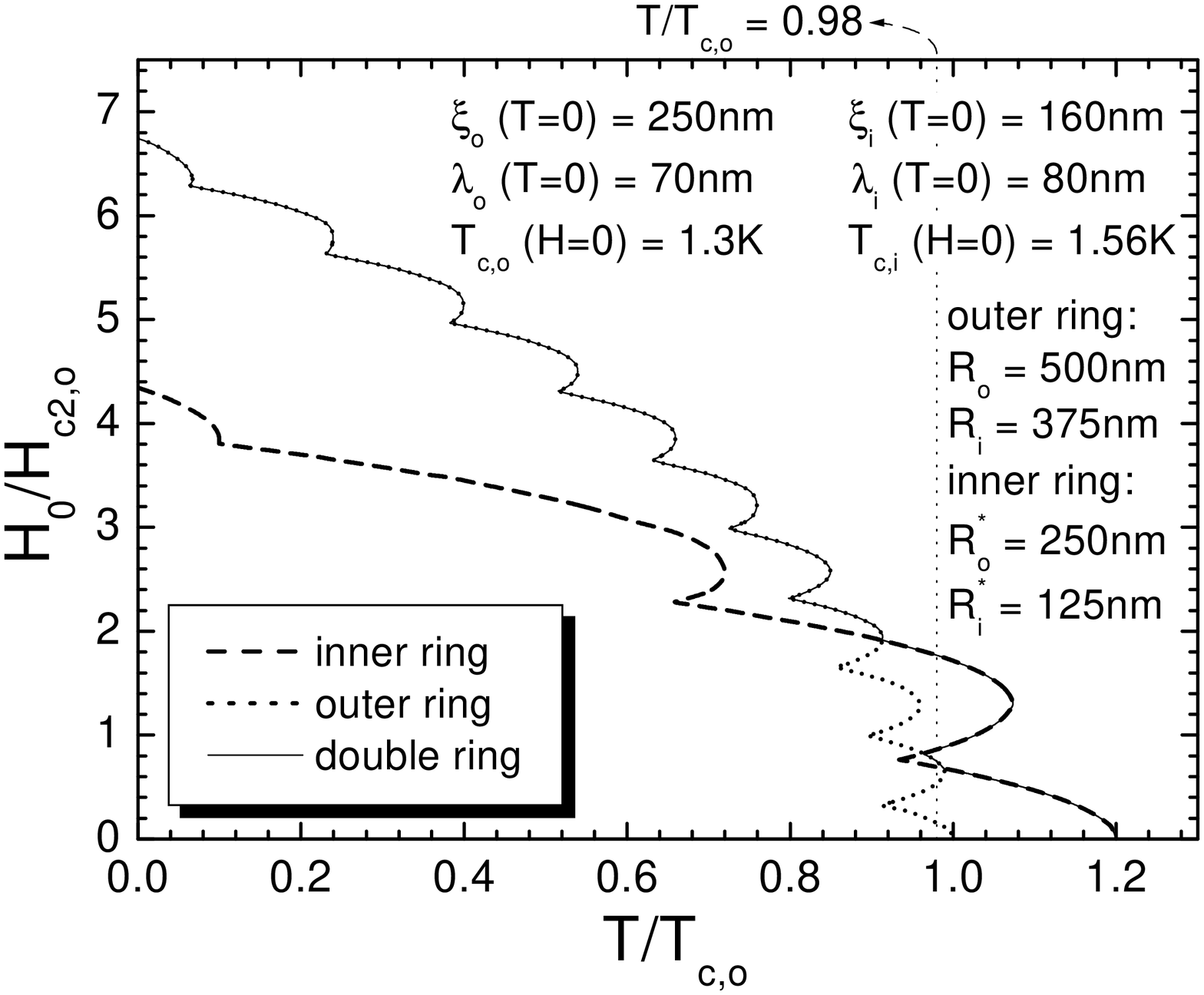, width=150mm, height=110mm, clip=}
\end{center}
\caption{The $H-T$ phase diagram for the inner ring (dashed curves), the
outer ring (dotted curves) and the double ring configuration (solid curves).
The material parameters and the sizes of both rings are different and are
given in the figure.}
\label{htphase}
\end{figure}

As an example, we take for the outer ring the values used by Geim {\it et al}%
~\cite{Geim1} for Al, i.e. $\xi _{o}\left( T=0\right) =250nm$, $\lambda
_{o}\left( T=0\right) =70nm$ and thus $\kappa _{o}=0.28$, resulting in a
critical temperature $T_{c,o}(H=0)=1.3K$. For the inner ring, we assume a
higher critical temperature $T_{c,i}=1.2T_{c,o}=1.56K$, and $\xi _{i}\left(
T=0\right) =160nm$, $\lambda _{i}\left( T=0\right) =80nm$ and thus $\kappa
_{i}=0.5$. For the radii of the outer ring we take as an example $%
R_{i}=375nm,$ $R_{o}=500nm,$ and for the inner ring $R_{i}^{\ast }=125nm$
and $R_{o}^{\ast }=250nm$. The $H-T$ phase diagram is shown in Fig.~\ref
{htphase} for the uncoupled situation for the inner ring (thick dashed
curves) and for the outer ring (thick dotted curves) and the coupled double
ring situation (solid curve). At $T=0$ the outer ring has a much higher
critical field $\left( H_{nuc}/H_{c2,o}=6.74\right) $ than the inner ring $%
\left( H_{nuc}/H_{c2,o}=4.34\right) $. Therefore, the superconducting/normal
transition of the double ring configuration equals the one of the outer ring
for low temperatures. With increasing temperature, the nucleation field of
the outer ring, i.e. the one of the double ring system, changes more quickly
than the one of the inner ring. The oscillations are the well known
Little-Parks oscillations. At $T/T_{c,o}=0.912$ both single rings have the
same transition field $H_{nuc}/H_{c2,o}=1.91$. At higher temperatures, the
superconducting/normal transition is determined by the inner ring.

\begin{figure}[tbp]
\begin{center}
\epsfig{file=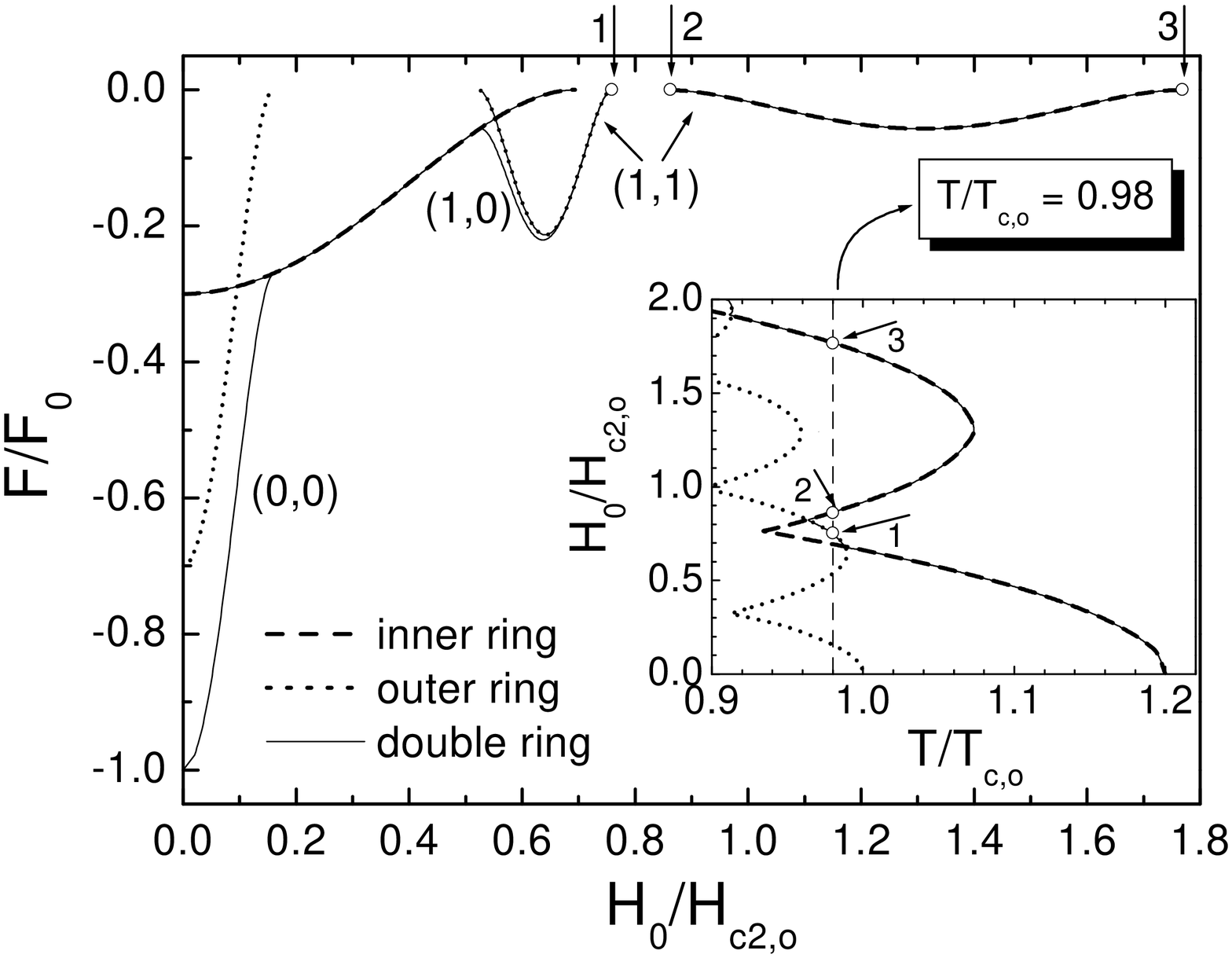, width=150mm, height=110mm, clip=}
\end{center}
\caption{The free energy as a function of the applied magnetic field for the
inner ring (dashed curves), the outer ring (dotted curves) and the double
ring configuration (solid curves) for the system of Fig.~\ref{htphase} at $%
T=0.98T_{c,o}$. The inset shows an enlargement of the phase diagram (Fig.~%
\ref{htphase}) in the $T/T_{c,o}\sim 1$\ region.}
\label{t098}
\end{figure}

The situation where the critical field of the outer ring is larger than the
one of the inner ring is exhaustively described in the previous sections. In
Fig.~\ref{t098} we show the free energy for the configuration of Fig.~\ref
{htphase} at $T=0.98T_{c,o}$ where the superconductivity of the inner ring
exists at larger fields than the one of the outer ring. The free energy of
the superconducting states of the inner ring are given by the dashed curves,
the states of the outer ring by the dotted curves and the double ring
configuration by the solid curves. Both in the inner and the outer ring,
superconducting states with vorticity $L=0$ and $L=1$ exist. At $%
T=0.98T_{c,o}$, the critical fields of the inner and the outer ring are $%
H_{0}/H_{c2,o}=1.77$ and $0.69$, respectively. Notice that in both rings the
free energies of the $L=0$ state and the $L=1$ state do not cross, which
means that with increasing field the ground state changes from the Meissner
state into the normal state and, with further increasing the field, into the 
$L=1$ state and back into the normal state. The reason is that near $T_{c}$
the superconductivity of the ring has decreased. This means that only rather
small currents can be induced and thus only a small flux can be attracted or
expelled by the ring. In the region between the existence of the $L=0$ state
and the $L=1$ state the currents, which have to be induced to expel or
attract the necessary flux to achieve vorticity $L=0$ and $L=1,$ are too
high. With increasing temperature, the $L=1$ state can not nucleate anymore
and the superconducting/normal transition jumps to the field where the $L=0$
state is destroyed. The corresponding oscillations in the $H-T$ phase
diagram\ (Fig.~\ref{htphase}) are the Little-Parks oscillations. For the
double ring configuration the $(0,0)$, the $(1,0)$ and the $(1,1)$ state can
nucleate. The $(1,1)$ state is split into two parts corresponding to the $%
L=1 $ states in the two single rings with an intermediate magnetic field
region in which both superconductors are normal. The ground state changes
from the Meissner state $(0,0)$ into the $(1,0)$ state at $%
H_{0}/H_{c2,o}\approx 0.53$, which equals the $(1,1)$ state at $%
H_{0}/H_{c2,o}>0.69$. Further increasing the field the ground state changes
into the normal state at $H_{0}/H_{c2,o}\approx 0.76$, then back into the $%
(1,1)$ state at $H_{0}/H_{c2,o}\approx 0.86$ and further back into the
normal state at $H_{0}/H_{c2,o}\approx 1.77$. Compared to the uncoupled
inner ring and the outer ring situation, extra ground state transitions
occur for the double ring case with interesting re-entrant superconducting
behavior and a switching on and off of the superconducting state in the
inner and outer ring.

\section{Two coupled {\it thin} rings in the limit $\protect\kappa \gg 1$}

Here we will show that in the limit of two coupled {\it thin} rings it is
possible to obtain analytical results for the coupling energy between the
two rings. This also corresponds to the case of $\kappa \gg 1$ and allows to
solve the problem analytically with the small parameter $d/\kappa ^{2}\ll 1$%
. From the numerical calculations of previous sections it follows that the
radial dependence of the order parameter in both inner and outer rings is
slow and smooth. Therefore, we assume that the order parameter in both inner
and outer rings, $f_{in}$ and $f_{out}$, respectively, are constant.\ In the
limit of the {\it thin} rings we neglect in the first approximation the $z-$%
dependence of the vector potential. With this assumption and because of the
cylindrical symmetry of the problem the vector potential has only the
azimuthal component $A\left( \rho \right) $ and the magnetic field has only
the normal component $H\left( \rho \right) =\partial \left( \rho A\left(
\rho \right) \right) /\partial \rho $.

The distribution of the vector potential due to the supercurrents inside the
inner and outer ring are described by the following equation (see also, for
example, Eqs.~(2)-(3) in Ref.~\cite{PRL81}): 
\begin{equation}
-\kappa _{i(o)}^{2}\frac{\partial }{\partial \rho }\left[ \frac{1}{\rho }%
\frac{\partial }{\partial \rho }\left( \rho A_{in(out)}\right) \right]
=d\left( \frac{L_{in(out)}}{\rho }-A_{in(out)}\right) f_{in(out)}^{2}, 
\eqnum{13a}  \label{VPring}
\end{equation}
and outside the rings by 
\begin{equation}
\frac{\partial }{\partial \rho }\left[ \frac{1}{\rho }\frac{\partial }{%
\partial \rho }\left( \rho A\right) \right] =0.  \eqnum{13b}  \label{VPspace}
\end{equation}
The corresponding boundary conditions are the continuity of both $A\left(
\rho \right) $\ and $H\left( \rho \right) $\ on the radial sides of the
rings. For $d\rightarrow 0$ it gives everywhere $A(d=0)=H_{0}\rho /2$ and $%
H(d=0)=H_{0}$. To a first approximation the solution of Eq.~(\ref{VPring})
becomes for small $d/\kappa _{i(o)}^{2}$%
\begin{equation}
A_{in(out)}\left( \rho \right) =\frac{H_{0}\rho }{2}+\frac{D_{in(out)}\rho }{%
2}+\frac{C_{in(out)}}{\rho }-\frac{d}{4\kappa _{i(o)}^{2}}%
f_{in(out)}^{2}\rho \left[ L_{in(out)}\left( 2\ln \rho -1\right) -H_{0}\rho
^{2}/4\right] ,  \eqnum{14a}  \label{VPSC}
\end{equation}
which is valid inside the superconductors, i.e. in the range $R_{i}^{\ast
}<\rho <R_{o}^{\ast }$\ and $R_{i}<\rho <R_{o}$.\ The vector potential
outside the rings is then obtain from a solution of Eq.~(\ref{VPspace}) and
is equal to 
\begin{equation}
A\left( \rho \right) =\left\{ 
\begin{array}{c}
D_{1}\rho /2,\qquad \qquad \qquad 0\leq \rho \leq R_{i}^{\ast }, \\ 
D_{3}\rho /2+C_{3}/\rho ,\qquad R_{o}^{\ast }\leq \rho \leq R_{i}, \\ 
H_{0}\rho /2+C_{ext}/\rho ,\qquad \qquad \rho \geq R_{o}.
\end{array}
\right.   \eqnum{14b}  \label{VPHoles}
\end{equation}
The corresponding distribution of the magnetic field is 
\begin{equation}
H\left( \rho \right) =\left\{ 
\begin{array}{c}
D_{1},\qquad \qquad \qquad \qquad \qquad \qquad \qquad \qquad \quad
\;\;\qquad 0\leq \rho \leq R_{i}^{\ast }, \\ 
H_{0}+D_{in}-f_{in}^{2}(d/\kappa _{i}^{2})\left( L_{in}\ln \rho -H_{0}\rho
^{2}/4\right) ,\quad \quad R_{i}^{\ast }\leq \rho \leq R_{o}^{\ast }, \\ 
D_{3},\qquad \qquad \qquad \qquad \qquad \qquad \qquad \qquad \quad \;\qquad
R_{o}^{\ast }\leq \rho \leq R_{i}, \\ 
H_{0}+D_{out}-f_{out}^{2}(d/\kappa _{o}^{2})\left[ L_{out}\ln \rho
-H_{0}\rho ^{2}/4\right] ,\quad R_{i}\leq \rho \leq R_{o}, \\ 
H_{0},\qquad \qquad \qquad \qquad \qquad \qquad \qquad \qquad \quad
\;\;\qquad \qquad \rho \geq R_{o}.
\end{array}
\right.   \eqnum{15}  \label{MF1}
\end{equation}
From the above boundary conditions the integration constants $D$\ and $C$\
in Eqs.~(\ref{VPSC})-(\ref{MF1}) are 
\begin{eqnarray*}
C_{in} &=&(d/4\kappa _{i}^{2})f_{in}^{2}R_{i}^{\ast 2}\left(
-L_{in}+H_{0}R_{i}^{\ast 2}/4\right) , \\
C_{3} &=&(d/4\kappa _{i}^{2})f_{in}^{2}\left( R_{o}^{\ast 2}-R_{i}^{\ast
2}\right) \left[ L_{in}-H_{0}\left( R_{o}^{\ast 2}+R_{i}^{\ast 2}\right) /4%
\right] , \\
C_{out} &=&C_{3}+(d/4\kappa _{o}^{2})f_{out}^{2}R_{i}^{2}\left(
-L_{out}+H_{0}R_{i}^{2}/4\right) , \\
C_{ext} &=&C_{3}+(d/4\kappa _{o}^{2})f_{out}^{2}\left(
R_{o}^{2}-R_{i}^{2}\right) \left[ L_{out}-H_{0}\left(
R_{o}^{2}+R_{i}^{2}\right) /4\right] , \\
D_{out} &=&(d/\kappa _{o}^{2})f_{out}^{2}\left[ L_{out}\ln
R_{o}-H_{0}R_{o}^{2}/4\right] , \\
D_{3} &=&H_{0}+(d/\kappa _{o}^{2})f_{out}^{2}\left[ L_{out}\ln \left(
R_{o}/R_{i}\right) -H_{0}S_{out}/4\pi \right] , \\
D_{in} &=&(d/\kappa _{i}^{2})f_{in}^{2}\left[ L_{in}\ln R_{o}^{\ast
}-H_{0}R_{o}^{\ast 2}/4\right] +(d/\kappa _{o}^{2})f_{out}^{2}\left[
L_{out}\ln \left( R_{o}/R_{i}\right) -H_{0}S_{out}/4\pi \right] , \\
D_{1} &=&H_{0}+(d/\kappa _{i}^{2})f_{in}^{2}\left[ L_{in}\ln \left(
R_{o}^{\ast }/R_{i}^{\ast }\right) -H_{0}S_{in}/4\pi \right] +(d/\kappa
_{o}^{2})f_{out}^{2}\left[ L_{out}\ln \left( R_{o}/R_{i}\right)
-H_{0}S_{out}/4\pi \right] .
\end{eqnarray*}
Notice that within this approximation the magnetic field is constant in the
inner hole and in the space between the two rings, and has a small (as the
ring's thickness is small) radial dependence inside the rings. Inserting the
expressions for the order parameters, the magnetic field and the vector
potential into the expression for the energy (note that in the equivalent
expression~(\ref{F_integr}) the Ginzburg-Landau equations have already\ been
used) 
\begin{equation}
F=\frac{2}{V}\int dV\left\{ -\frac{\xi _{o}^{2}}{\xi _{i,o}^{2}}\left| \Psi
\right| ^{2}+\frac{1}{2}\frac{\kappa _{i,o}^{2}}{\kappa _{o}^{2}}\left| \Psi
\right| ^{4}+\left| -i\overrightarrow{\nabla }\Psi -\overrightarrow{A}\Psi
\right| ^{2}+\kappa _{o}^{2}\left[ \overrightarrow{h}\left( \overrightarrow{r%
}\right) -\overrightarrow{H}_{0}\right] ^{2}\right\} .  \eqnum{16}
\label{EnergyGen}
\end{equation}
With a $O\left( d^{2}/\kappa ^{4}\right) $ accuracy we find the difference
between the Gibbs free energy of the superconducting state and the normal
state 
\begin{eqnarray}
F &=&\frac{2}{S}\left[ \left( -\frac{\xi _{o}^{2}}{\xi _{i}^{2}}f_{in}^{2}+%
\frac{1}{2}\frac{\kappa _{i}^{2}}{\kappa _{o}^{2}}f_{in}^{4}\right)
S_{in}+f_{in}^{2}\left( I_{in}^{(0)}-\left( d/\kappa _{i}^{2}\right)
f_{in}^{2}I_{in}^{\left( 1\right) }-\left( d/\kappa _{o}^{2}\right)
f_{out}^{2}I^{\left( 2\right) }\right) \right.   \eqnum{17}  \label{Energy1}
\\
&&\left. +\left( -f_{out}^{2}+\frac{1}{2}f_{out}^{4}\right)
S_{out}+f_{out}^{2}\left( I_{out}^{(0)}-\left( d/\kappa _{o}^{2}\right)
f_{out}^{2}I_{out}^{\left( 1\right) }-\left( d/\kappa _{i}^{2}\right)
f_{in}^{2}I^{\left( 2\right) }\right) \right] ,  \nonumber
\end{eqnarray}
where $S_{in}=\pi \left( R_{o}^{\ast 2}-R_{i}^{\ast 2}\right) $,$%
\;S_{out}=\pi \left( R_{o}^{2}-R_{i}^{2}\right) $, $S=S_{in}+S_{out},$ and 
\begin{eqnarray}
I_{out(in)}^{\left( 0\right) } &=&2\pi L_{out(in)}^{2}\ln \left(
R_{o}^{(\ast )}/R_{i}^{(\ast )}\right)
-H_{0}L_{out(in)}S_{out(in)}+H_{0}^{2}S_{out(in)}\left( R_{o}^{(\ast
)2}+R_{i}^{(\ast )2}\right) /8,  \eqnum{18}  \label{Integrals} \\
I^{(2)} &=&S_{in}\left[ L_{in}-H_{0}\left( R_{i}^{\ast 2}+R_{o}^{\ast
2}\right) /4\right] \left[ L_{out}\ln \left( R_{o}/R_{i}\right)
-H_{0}S_{out}/4\pi \right] ,  \nonumber \\
I_{out(in)}^{(1)} &=&2\left( L_{out(in)}\ln R_{o}^{(\ast
)}-H_{0}R_{o}^{(\ast )2}/4\right) J_{out(in)}^{(1)}-R_{i}^{(\ast )2}\left(
L_{out(in)}-H_{0}R_{i}^{(\ast )2}/4\right)
J_{out(in)}^{(2)}-J_{out(in)}^{(3)},  \nonumber \\
J_{out(in)}^{(1)} &=&\frac{S_{out(in)}}{2}\left[ L-H_{0}\left( R_{o}^{(\ast
)2}+R_{i}^{(\ast )2}\right) /4\right] ,  \nonumber \\
J_{out(in)}^{(2)} &=&\pi \left[ L_{out(in)}\ln \left( R_{o}^{(\ast
)}/R_{i}^{(\ast )}\right) -H_{0}S_{out(in)}/4\pi \right] ,  \nonumber \\
J_{out(in)}^{(3)} &=&\left. \pi \rho ^{2}\left[ L_{out(in)}^{2}\left( \ln
\rho -1\right) +L_{out(in)}H_{0}\rho ^{2}\left( 1-2\ln \rho \right)
/8+H_{0}^{2}\rho ^{4}/48\right] \right| _{R_{i}^{\left( \ast \right)
}}^{R_{o}^{\left( \ast \right) }}.  \nonumber
\end{eqnarray}

After the minimization of the free energy~(\ref{Energy1})\ with respect to $%
f_{in(out)}$ we obtain two equations 
\begin{equation}
f_{in(out)}\left[ \left( -\frac{\xi _{o}^{2}}{\xi _{i(o)}^{2}}+\frac{\kappa
_{i(o)}^{2}}{\kappa _{o}^{2}}f_{in(out)}^{2}\right)
S_{in(out)}+I_{in(out)}^{(0)}-\frac{2d}{\kappa _{i(o)}^{2}}%
f_{in(out)}^{2}I_{in(out)}^{(1)}-\left( \frac{d}{\kappa _{o}^{2}}+\frac{d}{%
\kappa _{i}^{2}}\right) f_{out(in)}^{2}I^{(2)}\right] =0,  \eqnum{19}
\label{EquilF}
\end{equation}
from which we obtain the equilibrium values of the order parameters for the
three possible situations.

When both the {\it inner }and{\it \ outer} rings are{\it \ }in the {\it %
superconducting state} (case I), we obtain with a $O\left( d^{2}/\kappa
^{4}\right) $ accuracy 
\begin{eqnarray}
\widetilde{f}_{in(out)}^{2} &=&\frac{\kappa _{o}^{2}}{\kappa _{i(o)}^{2}}%
\left( \frac{\xi _{o}^{2}}{\xi _{i(o)}^{2}}-\frac{I_{in(out)}^{(0)}}{%
S_{in(out)}}\right) \left( 1+\frac{2d}{\kappa _{i(o)}^{2}}\frac{\kappa
_{o}^{2}}{\kappa _{i(o)}^{2}}\frac{I_{in(out)}^{(1)}}{S_{in(out)}}\right)  
\eqnum{20}  \label{F2min1} \\
&&+\frac{d}{\kappa _{i}^{2}}\left( 1+\frac{\kappa _{o}^{2}}{\kappa _{i}^{2}}%
\right) \left( \frac{\xi _{o}^{2}}{\xi _{o(i)}^{2}}-\frac{I_{out(in)}^{(0)}}{%
S_{out(in)}}\right) \frac{I^{(2)}}{S_{in(out)}}.  \nonumber
\end{eqnarray}
Inserting these expressions into Eq.~(\ref{Energy1}) we obtain $%
F^{I}=F_{in}^{I}+F_{out}^{I}+F_{int}^{I}$, where 
\begin{equation}
F_{in}^{I}=-\frac{S_{in}}{S}\frac{\kappa _{o}^{4}}{\kappa _{i}^{4}}\left( 
\frac{\xi _{o}^{2}}{\xi _{i}^{2}}-\frac{I_{in}^{(0)}}{S_{in}}\right) ^{2}%
\left[ \frac{\kappa _{i}^{2}}{\kappa _{o}^{2}}+\frac{2d}{\kappa _{i}^{2}}%
\frac{I_{in}^{(1)}}{S_{in}}+O\left( \frac{d^{2}}{\kappa _{i}^{4}}\right) %
\right] ,  \eqnum{21a}  \label{G_in1}
\end{equation}
is the self energy of the inner ring, and 
\begin{equation}
F_{out}^{I}=-\frac{S_{out}}{S}\left( 1-\frac{I_{out}^{(0)}}{S_{out}}\right)
^{2}\left[ 1+\frac{2d}{\kappa _{o}^{2}}\frac{I_{out}^{(1)}}{S_{out}}+O\left( 
\frac{d^{2}}{\kappa _{o}^{4}}\right) \right] ,  \eqnum{21b}  \label{G_out1}
\end{equation}
is the self energy of the outer ring, while 
\begin{equation}
F_{int}^{I}=-\frac{2d}{\kappa _{i}^{2}}\left( 1+\frac{\kappa _{o}^{2}}{%
\kappa _{i}^{2}}\right) \left( \frac{\xi _{o}^{2}}{\xi _{i}^{2}}-\frac{%
I_{in}^{(0)}}{S_{in}}\right) \left( 1-\frac{I_{out}^{(0)}}{S_{out}}\right) 
\frac{I^{(2)}}{S}+O\left( \frac{d^{2}}{\kappa ^{4}}\right) ,  \eqnum{21c}
\label{G_int1}
\end{equation}
is the interaction energy between the two rings.

When one of the rings is in the normal state the results become more simple.
Namely, when the outer ring is in the{\it \ }normal state and the{\it \ inner%
} ring is {\it superconducting} (case~II) or vice verca (case~III), we have 
\begin{equation}
\widetilde{f}_{out}=0,\quad \widetilde{f}_{in}^{2}=\left( \frac{\xi _{o}^{2}%
}{\xi _{i}^{2}}-\frac{I_{in}^{(0)}}{S_{in}}\right) \left( \frac{\kappa
_{i}^{2}}{\kappa _{o}^{2}}-\frac{2d}{\kappa _{i}^{2}}\frac{I_{in}^{(1)}}{%
S_{in}}\right) ^{-1},  \eqnum{22}  \label{F2min2}
\end{equation}
and 
\begin{equation}
\widetilde{f}_{in}=0,\quad \widetilde{f}_{out}^{2}=\left( 1-\frac{%
I_{out}^{(0)}}{S_{out}}\right) \left( 1-\frac{2d}{\kappa _{o}^{2}}\frac{%
I_{out}^{(1)}}{S_{out}}\right) ^{-1},  \eqnum{23}  \label{F2min3}
\end{equation}
respectively. The corresponding energies are 
\begin{equation}
F^{II}=-\frac{S_{in}}{S}\left( \frac{\xi _{o}^{2}}{\xi _{i}^{2}}-\frac{%
I_{in}^{(0)}}{S_{in}}\right) ^{2}\left( \frac{\kappa _{i}^{2}}{\kappa
_{o}^{2}}-\frac{2d}{\kappa _{i}^{2}}\frac{I_{in}^{(1)}}{S_{in}}\right) ^{-1},
\eqnum{24}  \label{EnFmin2}
\end{equation}
\begin{equation}
F^{III}=-\frac{S_{out}}{S}\left( 1-\frac{I_{out}^{(0)}}{S_{out}}\right)
^{2}\left( 1-\frac{2d}{\kappa _{o}^{2}}\frac{I_{out}^{(1)}}{S_{out}}\right)
^{-1},  \eqnum{25}  \label{EnFmin3}
\end{equation}
and in the $d/\kappa _{i(o)}^{2}\ll 1$\ limit they coincide with $F_{in}^{I}$%
\ and $F_{out}^{I}$, respectively.

One can see, that an interaction between the two rings (i.e., the coupling)
exists only when {\it both rings are superconducting.} The energy of the
ring-ring coupling in the considered limit is proportional to the ring's
thickness. Due to the interaction between the rings the Cooper pair density
in each ring has a small (proportional to $d/\kappa ^{2}$) contribution from
the neighbouring ring.

\begin{figure}[tbp]
\begin{center}
\epsfig{file=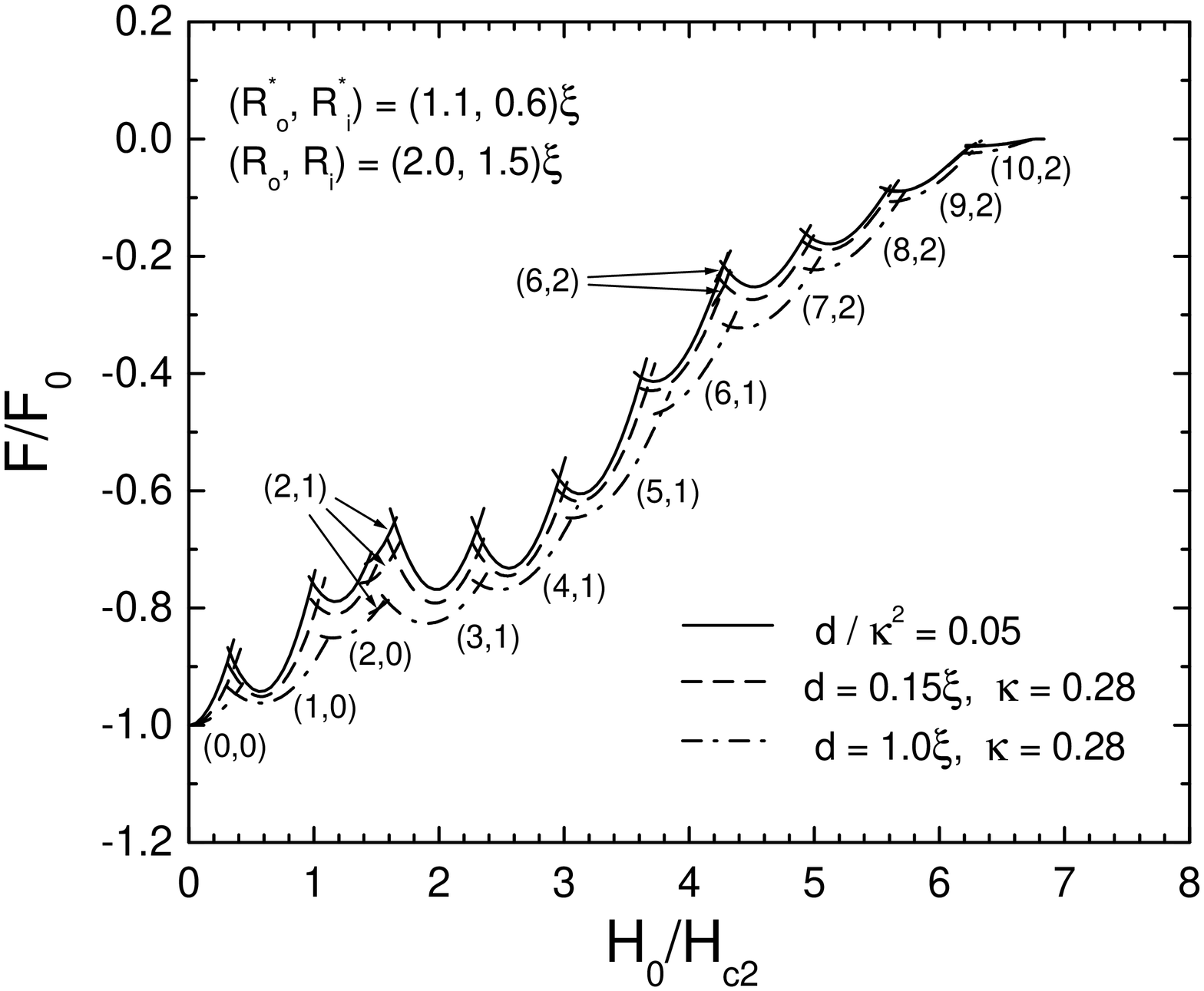, width=150mm, height=110mm, clip=}
\end{center}
\caption{The groud state free energy of the double ring configuration with
the same parameters as in Fig.~\ref{emaggrrr} obtained from our analytical
expressions
which are valid for type-II superconducting rings. Also the ground state
energies from Fig.~\ref{emaggrrr} are shown.}
\label{Figtheor}
\end{figure}

In Fig.~\ref{Figtheor} the calculated magnetic field dependence of the
ground state energy of the coupled thin rings of the same material (i.e.,
with $\xi _{i}=\xi _{o}$ and $\kappa _{i}=\kappa _{o}$)\ with the same
radial sizes as in Fig.~5 is shown for $d/\kappa ^{2}=0.05$ (solid curve).
Also for comparison the curves from Fig.~5 are shown by dashed and
dash-dotted lines which correspond to $d/\kappa ^{2}\simeq 1.98$ and $12.76$%
, respectively. One can see that all curves have the same qualitative
behaviour and with increasing $d/\kappa ^{2}$ only a small decrease of the
ground state energy takes place. Also the values of the transition fields
between the different $L$ states are very nicely reproduced. This is quite
surprising and, to make the physics of this result more clear, we consider 
{\it the limit of narrow rings}.

Let's introduce the average radius of the two rings, $\overline{\rho }%
_{in}=\left( R_{o}^{\ast }+R_{i}^{\ast }\right) /2,\overline{\rho }%
_{out}=\left( R_{o}+R_{i}\right) /2$, respectively, and their corresponding
widths, $2w_{in}=R_{o}^{\ast }-R_{i}^{\ast },2w_{out}=R_{o}-R_{i}$. Next we
expand Eqs.~(\ref{Integrals}) and~(21)\ with respect to the small parameters 
$2w_{in(out)}/\overline{\rho }_{in(out)}\ll 1$ with an $O\left(
w_{in(out)}^{2}\right) $ accuracy. Within this approximation the
energies~(21) are 
\begin{eqnarray}
F_{in(out)}^{I} &=&-\frac{S_{in(out)}}{S}\frac{\kappa _{o(i)}^{2}}{\kappa
_{i}^{2}}\left[ \frac{\xi _{o(i)}^{2}}{\xi _{i}^{2}}-\frac{\left(
L_{in(out)}-\overline{\Phi }_{in(out)}\right) ^{2}}{\overline{\rho }%
_{in(out)}^{2}}\right]   \eqnum{26a}  \label{Finoutasympt} \\
&&\times \left[ 1+\frac{2d}{\kappa _{i(o)}^{2}}\frac{\kappa _{o(i)}^{2}}{%
\kappa _{i}^{2}}\frac{2w_{in(out)}}{\overline{\rho }_{in(out)}}\left(
L_{in(out)}-\overline{\Phi }_{in(out)}\right) ^{2}+O\left( \frac{d}{\kappa
_{i(o)}^{2}}w_{in(out)}^{2}\right) \right]   \nonumber \\
&&\times \left[ \frac{\xi _{o(i)}^{2}}{\xi _{i}^{2}}-\frac{\left(
L_{in(out)}-\overline{\Phi }_{in(out)}\right) ^{2}}{\overline{\rho }%
_{in(out)}^{2}}+O\left( w_{in(out)}^{2}\right) \right] ,  \nonumber
\end{eqnarray}
\begin{eqnarray}
F_{int}^{I} &=&-\frac{S_{in}}{S}\frac{2d}{\kappa _{i}^{2}}\left\{ \frac{%
2w_{out}}{\overline{\rho }_{out}}\left( 1+\frac{\kappa _{o}^{2}}{\kappa
_{i}^{2}}\right) \left( L_{in}-\overline{\Phi }_{in}\right) \left( L_{out}-%
\overline{\Phi }_{out}\right) \right.   \eqnum{26b}  \label{Fintasympt} \\
&&\left. \times \left[ \frac{\xi _{o}^{2}}{\xi _{i}^{2}}-\frac{\left( L_{in}-%
\overline{\Phi }_{in}\right) ^{2}}{\overline{\rho }_{in}^{2}}\right] \left[
1-\frac{\left( L_{out}-\overline{\Phi }_{out}\right) ^{2}}{\overline{\rho }%
_{out}^{2}}\right] +O\left( w_{in}w_{out},w_{out}^{2}\right) \right\} , 
\nonumber
\end{eqnarray}
where $\overline{\Phi }_{in(out)}=H_{0}\overline{\rho }_{in(out)}^{2}/2$ is
the average magnetic flux through the corresponding ring, measured in units
of $\Phi _{0}=hc/2e$. From these expressions one can see that the dominant
thickness dependent terms in $F_{in(out)}^{I}$, as well as the ring's
interaction energy $F_{int}^{I}$, are of order $O\left( \left( d/\kappa
_{i(o)}^{2}\right) w_{in(out)}\right) $ (not of order $O\left( d/\kappa
_{i(o)}^{2}\right) $, as one could think naively). This additional smallness
shows that an increase of the ring thickness influences the ring's
self-energies and the interaction energy very slightly and for rings which
have a not too large width (as in our case) the above results become valid
for $d/\kappa _{i(o)}^{2}\gtrsim 1$.

\section{Conclusions}

We investigated the magnetic coupling between two concentric mesoscopic
superconductors with non-zero thickness. When a second superconductor is
placed in the center of a superconducting ring, it feels a non-uniform
field, which is the superposition of the uniform applied field and the field
expelled from the outer ring. Also the first ring will be influenced by the
magnetic field expelled from the superconductor in the center. So, both
superconductors are coupled magnetically. This results in substantial
changes of the superconducting properties.

From the study of the free energy we learned that extra ground state
transitions occur in comparison with the single ring case. These are
transitions where the total vorticity stays the same, but the vorticity of
the inner superconductor changes by one unit. We also found that the free
energy of the double ring system is not exactly the same as the sum of the
free energies of the two uncoupled single rings which is another signature
of the magnetic coupling of both rings. This interaction enhances with
increasing sample thickness. We also calculated the expelled field for the
ring-ring configuration which showed that as compared with a single ring
more, or less, field can be expelled or attracted depending on the
vorticities of both superconductors.

The behaviour of the Cooper-pair density, the magnetic field profile and the
current density was calculated. Since an extra superconductor is placed in
the center, the magnetic field will be expelled from this superconductor or
will be compressed in the center of it, which results in a higher or a lower
magnetic field density between the two superconductors. The current in both
rings exhibits extra jumps at the transition fields where the vorticity of
the other ring increases or decreases by one. The reason is that at these
applied fields the total magnetic field in the region between the two
superconductors changes.

We investigated what happens if the inner ring is made of a different
material with a higher critical temperature. The $H-T$ phase diagram showed
that the nucleation field of the double ring equals the one of the outer
ring at low temperatures and the one of the inner ring at higher
temperatures.

Analytical expressions are obtained for the magnetic field distribution and
the energy of two coupled {\it thin} type-II superconducting rings. These
analytical results are found to give excellent results when $d/\kappa
_{i(o)}^{2}<1$ and, moreover, give also good agreement with our ``exact''
numerical results for sufficiently narrow rings when $d/\kappa _{i(o)}^{2}>1$%
.

\section{Acknowledgments}

This work was supported by the Flemish Science Foundation (FWO-Vl), the
''Onderzoeksraad van de Universiteit Antwerpen'' (GOA), the
''Interuniversity Poles of Attraction Program - Belgian State, Prime
Minister's Office - Federal Office for Scientific, Technical and Cultural
Affairs'', and the ESF ``VORTEX'' Programme. Discussions with
V.~V.~Moshchalkov and M. Morelle are gratefully acknowledged.

\bigskip


\end{document}